\documentclass[twocolumn,
amsmath,
amssymb,
amsfonts,
final,
a4paper,
10pt,
aps,
twocolumn,
citeautoscript,
footinbib,
pra,
superscriptaddress]{revtex4-2}
\usepackage{times}
\usepackage{graphicx}
\usepackage[latin1]{inputenc}
\usepackage[T1]{fontenc}
\usepackage[
   breaklinks=true,
   colorlinks=true,
   bookmarks=true,
   bookmarksopenlevel=3, 
   bookmarksnumbered=true,
   menucolor=black,
   linkcolor=blue,
   citecolor=blue,
   urlcolor=blue,
   pdftitle={},
   pdfauthor={Guillaume Roux},
   pdfcreator={GhostScript},
   pdfsubject={},
   pdfkeywords={},
   pdfstartview=FitH]{hyperref}

\usepackage{physics}

\newcommand{\XXZ}{XXZ}
\newcommand{\hsigma}{\hat{\sigma}}
\newcommand{\hSigma}{\hat{\Sigma}}
\newcommand{\Htilde}{\hat{H}^{(eg)}}   
\newcommand{\ha}{\hat{a}}
\newcommand{\polaronH}{\hat{\widetilde{H}}}

\usepackage{siunitx}
\DeclareSIUnit\Gauss{Gauss}
\begin{document}

\author{Paul \surname{M\'ehaignerie}}
\affiliation{Laboratoire Kastler Brossel, Coll\`ege de France,
 CNRS, ENS-Universit\'e PSL,
 Sorbonne Universit\'e, \\11, place Marcelin Berthelot, 75005 Paris, France}
\author{Cl\'ement \surname{Sayrin}}
\affiliation{Laboratoire Kastler Brossel, Coll\`ege de France,
 CNRS, ENS-Universit\'e PSL,
 Sorbonne Universit\'e, \\11, place Marcelin Berthelot, 75005 Paris, France}
\author{Jean-Michel \surname{Raimond}}
\affiliation{Laboratoire Kastler Brossel, Coll\`ege de France,
 CNRS, ENS-Universit\'e PSL,
 Sorbonne Universit\'e, \\11, place Marcelin Berthelot, 75005 Paris, France}
\author{Michel \surname{Brune}}
\affiliation{Laboratoire Kastler Brossel, Coll\`ege de France,
 CNRS, ENS-Universit\'e PSL,
 Sorbonne Universit\'e, \\11, place Marcelin Berthelot, 75005 Paris, France}
\author{Guillaume \surname{Roux}}
\email{guillaume.roux@universite-paris-saclay.fr}
\affiliation{Universit\'e Paris-Saclay, CNRS, LPTMS, 91405, Orsay, France.}

\date{\today} 

\title{Spin-motion coupling in a circular Rydberg state quantum simulator: case of two atoms}


\begin{abstract}

Rydberg atoms are remarkable tools for the quantum simulation of spin arrays. Circular Rydberg atoms open the way to simulations over very long time scales, using a combination of laser trapping of the atoms and spontaneous-emission inhibition, as shown in the proposal of a \XXZ\ spin-array simulator based on chains of trapped circular atoms  [T.L. Nguyen \textit{et al.}, Phys. Rev. X \textbf{8}, 011032
(2018)]. Such simulators could reach regimes (thermalization, glassy dynamics) that are out of the reach of those based on ordinary, low-angular-momentum short-lived Rydberg atoms. Over the promised long time scales, the unavoidable coupling of the spin dynamics with the atomic motion in the traps may play an important role. We study here the interplay between the spin exchange and motional dynamics in the simple case of two interacting circular Rydberg atoms confined in harmonic traps. 
The time evolution is solved exactly when the position dependence of the dipole-dipole interaction terms can be linearized over the extension of the atomic motion. We present numerical simulations in more complex cases, using the realistic parameters of the simulator proposal. 
We discuss three applications. 
First, we show that realistic experimental parameters lead to a regime in which atomic and spin dynamics become fully entangled, generating interesting non-classical motional states. 
We also show that, in other parameter regions, the spin dynamics notably depends on the initial temperature of the atoms in the trap, providing a sensitive motional thermometry method. Last, and most importantly, we discuss the range of parameters in which the motion has negligible influence over the spin dynamics.
\end{abstract}

\maketitle 

\section{Introduction}

Quantum simulation is a particularly important and active field of the thriving quantum technologies~\cite{PreskillQuantumComputingNISQ2018,AltmanQuantumSimulatorsArchitectures2021}. It opens the way to significant advances in the understanding and design of quantum materials, in quantum chemistry and, more generally in complex optimization problems. Many platforms are actively considered to implement quantum simulators, including superconducting qubits architectures~\cite{ZhuObservationThermalizationInformation2022,MiTimecrystallineeigenstateorder2022}, trapped ions~\cite{MonroeProgrammablequantumsimulations2021}, neutral atoms in optical lattices~\cite{HofstetterQuantumsimulationstrongly2018,WeiQuantumgasmicroscopy2022} and Rydberg atoms~\cite{SchaussQuantumsimulationtransverse2018,BrowaeysManybodyphysicsindividually2020}.

The extremely strong interaction between Rydberg atoms (highly excited states with a principal quantum number, $n$, typically of the order of 50) is a particularly promising tool for quantum simulation. The dipole blockade mechanism acting on the laser excitation toward Rydberg levels from ground-state atoms trapped in programmable arrays of optical tweezers already led to simulations of spin arrays in conditions where exact numerical approaches with classical computing devices are difficult or impossible. Notable milestones are the observations of quantum antiferromagnetic correlations~\cite{EbadiQuantumphasesmatter2021,SchollQuantumsimulation2D2021}, of quantum phase transitions~\cite{KeeslingQuantumKibbleZurek2019}, of spin liquids~\cite{SemeghiniProbingtopologicalspin2021a} and of dynamics related to quantum scars~\cite{BluvsteinControllingquantummanybody2021}. 

Most of the Rydberg quantum simulation experiments so far have been based on laser-accessible, low-angular momentum, $\ell$,  states. These states have  at most a few hundred microseconds lifetime, dominated by direct optical transitions toward low-lying levels. The effective simulation time for an array made up of a few hundred atoms is thus at most in the $\si{\micro\second}$ range, corresponding to a few cycles of the spin flip-flop between nearest neighbors. On the one hand, over such short times, atomic motion is  irrelevant (be it due to the finite atomic temperature in the tweezers or to the dipolar interaction forces themselves). On the other hand, this fundamental limitation of the simulation time makes it difficult to study some interesting phenomena, such as thermalization or disorder-induced slow glassy dynamics.

\begin{figure}[t]
\centering
\includegraphics[width=0.95\columnwidth]{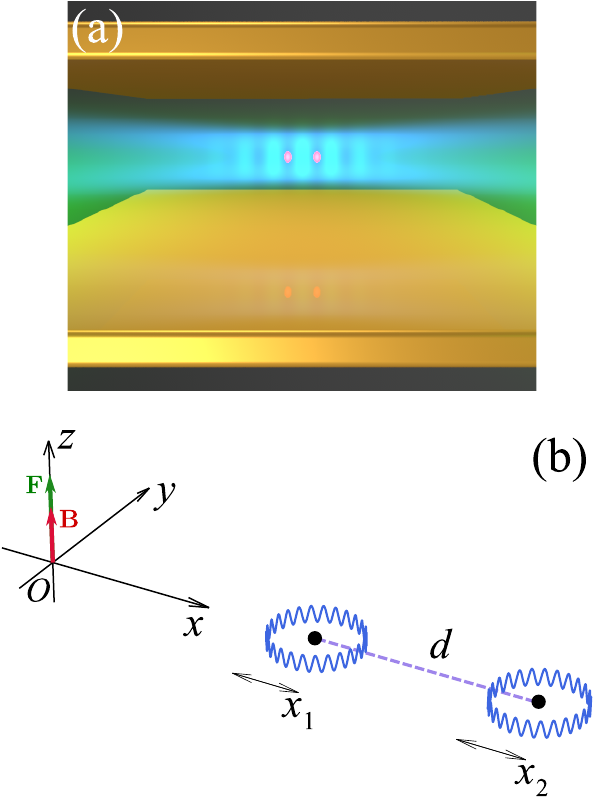}
\caption{(a) Pictorial scheme of the proposed circular state quantum simulator with two interacting Rydberg atoms.  Adapted from ~\cite{NguyenQuantumSimulationCircular2018}. (b) Sketch with notations and respective orientations.}
\label{fig:picscheme}
\end{figure}

To circumvent this limitation, we have recently proposed a quantum simulator based on circular Rydberg atoms~\cite{NguyenQuantumSimulationCircular2018}. These levels have maximum angular and magnetic quantum numbers ($|m|=\ell=n-1$). Their spontaneous-emission lifetime ($\SI{30}{\milli\second}$ for $n=50$) is much longer than that of ordinary Rydberg atoms, since the only decay channel is a $\sigma^+$-polarized millimeter-wave transition. Blackbody-radiation (BBR) induced transfers on the microwave transitions out of the circular state $nC$ are efficiently suppressed in a cryogenic environment. Lifetimes of up to $\SI{10}{\milli\second}$ have already been observed with a radiation environment at $\SI{10}{\kelvin}$~\cite{Cantat-MoltrechtLonglivedcircularRydberg2020}. This long lifetime can be considerably lengthened by placing the circular atoms in a plane-parallel capacitor inhibiting their main $\sigma^+$-polarized spontaneous or BBR-induced decay channels [see Fig.~\ref{fig:picscheme}(a)]. Conservative estimates indicate that the individual atomic lifetime could reach the minute range~\cite{NguyenQuantumSimulationCircular2018}. 

Taking benefit of such long lifetimes makes it necessary to laser-trap the atoms, e.g., in the ponderomotive potential experienced by their nearly free outer electron in a laser beam. Defect-free atomic chains with tens of atoms trapped in a linear array of laser traps can be prepared by an original evaporation mechanism  \cite{NguyenQuantumSimulationCircular2018,BruneEvaporativecoolingRydberg2020}. Arrays of hundred of circular atoms held in adapted optical tweezers could also be prepared using well-established techniques~\cite{BarredoSyntheticthreedimensionalatomic2018,BarredoThreeDimensionalTrappingIndividual2020}. The interactions between nearest-neighbor circular atoms naturally implement the Hamiltonian of a \XXZ\ spin array. Interestingly, all parameters of the Hamiltonian can be tuned over a wide range by adjusting the electric, magnetic and microwave-dressing fields applied onto the atoms. Circular atoms quantum simulators thus open the way to an entirely new simulation regime.

Over such long timescales, atomic motion in the laser traps cannot be a priori neglected. In particular, the spin exchange flip-flops are bound to create a modulation of the van der Waals forces between adjacent atoms. This may lead to a significant heating of the atomic array or even to the full entanglement of the spin and motional dynamics. Interestingly, such coupling between electronic states and vibrational modes has recently been proposed to engineer multi-body interactions~\cite{Gambetta2020} and mimick molecular dynamics~\cite{Magoni2023}.
The dressing of spin excitations by phonon modes leads to the formation of polaronic quasiparticle excitation, which dispersion relation has been discussed in Refs.~\onlinecite{Mazza2020, Magoni2022}. Note that these studies discussed a situation different from that considered here~\cite{NguyenQuantumSimulationCircular2018}, since the two spin states  then correspond to the ground state and to a laser-accessible Rydberg state, with an interaction based on the dipole blockade or facilitation mechanisms.
Last, protocols decoupling electronic and motion dynamics have been addressed in Ref.~\onlinecite{Cohen2021} in a related configuration.

In Ref.~\onlinecite{NguyenQuantumSimulationCircular2018}, we have shown by qualitative arguments that, for selected and realistic operational parameters of the proposed simulator, the spin-motion entanglement can be made negligible as compared to the \XXZ\ spin Hamiltonian dynamics. Here, we plan to substantiate this claim by a detailed analysis of the motion along the interatomic axis of two adjacent coupled atoms in a circular-state simulator. To unveil the main mechanisms, we focus on the simple dynamics induced by spin excitation exchange between two atoms. Our approach is based on analytic solutions obtained at the expense of well-controlled approximations and on numerical simulations of the exact system.

The paper is organized as follows. Section II describes in more details the trapping potential and introduces the relevant degrees of freedom, the interaction potentials and the  resulting spin-motion  Hamiltonian with realistic experimental parameters.  Section III introduces a simple and global picture for the dynamics in terms of interfering paths and an oscillator-spin coupling picture. Section IV shows that a linear approximation of the spin-motion Hamiltonian, valid for small enough atomic displacements, is analytically tractable. It allows us to provide quantitative predictions for the
the creation of motional non-classical ``cat states'' in a strong-interaction and weak-trap limit. 
In Section V, we extend the results to finite atomic temperatures.
The initial motion of the atoms due to their finite temperature comes into play and gradually reduces the spin exchange oscillation contrast. We show that this could lead, interestingly, to a precise thermometry of the atomic motion.
Turning to the simulator regime in Section VI, we show that, for strong enough laser trapping potentials, the spin  dynamics is effectively dominant over long timescales, of the order of hundreds of spin-exchange times. 
We provide analytic predictions for the effects of second order terms, which capture the non-linearities in this regime. Two appendices provide a discussion of the stability of the trapping potential for very weak traps (Appendix \ref{app:stability}) and some useful analytical formulae (Appendix \ref{app:formulas}).

\section{Model and parameters}

\subsection{Experimental context}

We discuss here the spin/motion coupling in the context of the circular-state simulator proposal described in \cite{NguyenQuantumSimulationCircular2018}. Its principle is shown in Fig.~\ref{fig:picscheme}. The circular atoms are placed in a plane parallel capacitor, which plays two roles. On the one hand, it provides a directing electric field $\mathbf{F}$ aligned along the $(Oz)$ quantization axis (axes definition in Fig.~\ref{fig:picscheme}). The nearly planar orbit of the circular atoms is thus parallel to the capacitor plates. A few Gauss homogeneous magnetic field $\mathbf{B}$, aligned with $(Oz)$, too, is used to tune the interatomic interaction Hamiltonian. On the other hand, the capacitor inhibits the $\sigma^+$-polarized spontaneous emission and BBR-induced transfers out of the circular states.

The circular atoms are trapped by a laser-induced ponderomotive potential acting on their nearly free outer electron.
For the sake of definiteness, we will base our discussions on the parameters of the ponderomotive trap for a chain of atoms described in~ \cite{NguyenQuantumSimulationCircular2018}. Our results can nevertheless be applied, \textsl{mutatis mutandis}, to all trap geometries, provided that the motion can be considered unidimensional along the interatomic axis. Note that this is not a stringent hypothesis, since transverse motion modifies the interatomic distance, determining the strength of the dipole-dipole interactions, only to second order. 

The positive ponderomotive energy, proportional to the laser intensity $I$, confines the atoms near an intensity minimum. Periodic minima along the $(Ox)$ interatomic axis are created by two non-interfering laser beams arrangements at a wavelength close to $\SI{1}{\micro\meter}$. A hollow Laguerre-Gauss $LG_{0,1}$mode  provides a tight trapping in the radial direction. Two coherent Gaussian beams interfering at a shallow angle in the $(xOy)$ plane create a periodic sine potential along the $(Ox)$ axis, with an adjustable spacing $d$ between minima, determining the spacing of the final spin array. In this geometry, the interatomic axis ($Ox$) is  perpendicular to the quantization one ($Oz$). An evaporative cooling procedure \cite{NguyenQuantumSimulationCircular2018,BruneEvaporativecoolingRydberg2020} prepares with a high probability defect-free chains with up to $N\simeq 100$ atoms. Here, we consider only the case of the simplest non-trivial chain with $N=2$ atoms. Generalization to longer chains will be the subject of further work.

With experimentally realizable powers, the transverse trapping angular frequency, $\omega_\perp$, can be made much larger than the longitudinal trapping frequency, $\omega$. We can thus neglect the transverse motion and consider only a unidimensional atomic motion along $(Ox)$ around the equilibrium positions. Provided that the extension of the atomic motion is small as compared to the average distance to the longitudinal potential maximum ($d/2$) or, equivalently, that the atomic motional energy is small as compared to the longitudinal trap depth (a few MHz typically), we can consider the trapping potential as harmonic.

\begin{table*}[t]
\begin{tabular}{|c||c|c|c|c|c||c|c|c|c||c|c|c||c|c||c|c|c|c|c|c|}
\hline
Type & $n$,$\Delta n$& $m$  & $d$  & $\frac{\omega}{2\pi}$ & $T_h$  & $\frac{J}{2\pi}$  & $\frac{J_z}{2\pi}$  & $\frac{\delta E_0}{2\pi}$  & $\frac{\delta \zeta}{2\pi}$  & $J_z/J$ & $\Delta_0/J$ & $\omega/J$ & $x_0$  & $|\Omega|/2\pi$ & $g$ & $\alpha$ & $\alpha_z$  \\
& & & [$\si{\micro\meter}$] &  [kHz] &  [$\si{\micro\kelvin}$] &  [kHz] &  [kHz] & [kHz] &  [kHz] & & &  &   [nm] &  [kHz] &  &  &   \\
\hline
\hline
\textbf{ XXZ} & $48, 2$ & $6$ &   $6$ & $50$ & $2.4$ & $-5.75$ & $-6.10$ & $67.1$ & $-9.78$ & $1.06$ &  $12.7$ & $-8.7$ & $48.2$ & $22.83$ & $0.008$ & $-0.011$ & $-0.071$ \\
 \hline
\textbf{THM} & $48, 2$ & $6$ &  $6$ & $15$ & $0.72$  & $-5.75$ &$-6.10$ & $67.1$ & $-9.78$ & $1.06$ &  $12.7$ & $-2.61$ & $87.7$ & $21.25$ & $0.0146$ & $-0.067$ & $-0.43$\\
\hline
\textbf{CATS} & $48, 1$ &$ 3$ &   $8$ & $50$ & $2.4$  & $-1277$ & $-9.49$ & 20 & $-0.87$ & $0.0074$ & $0.023$ & $0.032$ & $48.2$ & $5105$ & $0.006$ & $-0.924$ & $-0.0141$\\

\hline
\end{tabular}
\caption{Parameters of the Hamiltonian for the three cases discussed in this paper (all the parameters are defined in the main text). Note that for the \textbf{CATS} set of parameters, the  $1/d^3$ dependence only applies to the $J$ spin-exchange coupling, while $J_z$, $\delta E_0$ and $\delta \zeta$  vary with the interatomic distance as $1/d^6$. The latter are therefore taken into account only in the exact numerical approach, while they are set to zero in the analytical formulae pertaining to this case.}
\label{tab:vdw}
\end{table*}

\subsection{Spin-exchange Hamiltonian}

In Ref.~\onlinecite{NguyenQuantumSimulationCircular2018}, the spin states in the simulated model are represented by two circular states $\{\ket{g}=\ket{nC},\ket{e}=\ket{(n+\Delta n)C}\}$, with $\Delta n=2$. For numerical evaluations, we will consider the case $n=48$ throughout this paper. We use the notation $\hat{\sigma}^{X,Y,Z}$ for the Pauli matrices in the $\{\ket g, \ket e\}$ basis. The dipole-dipole interaction then acts at the second order only in a perturbation expansion, both for the interaction between two atoms in the same state and for the spin-exchange process. Assuming, for the time being, the atoms to be at rest at a relative distance $d$, the spin-spin interaction Hamiltonian  reads~\cite{NguyenQuantumSimulationCircular2018}:
\begin{align}
\frac{\hat{H}_S}{\hbar} &=  J_z \hsigma_1^Z \hsigma_2^Z 
+ J \left( \hsigma_1^X \hsigma_2^X + \hsigma_1^Y \hsigma_2^Y \right)\nonumber\\
&\quad +\delta E_0  + \frac{\delta \zeta}{2} \left(\hsigma_1^Z + \hsigma_2^Z \right)\;,
\label{eq:HS}
\end{align}
where the indices refer to the atom number (see Fig.~\ref{fig:picscheme}). For $d=\SI{6}{\micro\meter}$,  for instance, the spin exchange term is  $J=-2\pi\times \SI{5.75}{\kHz}$. The spin-spin term, $J_z$, depends upon the electric and magnetic fields applied on the atoms. We consider here the typical situation with $F=\SI{9}{\volt\per\centi\meter}$ and $B=\SI{12}{\Gauss}$, corresponding to $J_z=-2\pi\times 6.2$~kHz.  This choice, leading to  $J_z\simeq J$, corresponds, in terms of the \XXZ\ model phase diagram, to the interesting transition between the Luttinger liquid and N\'eel order phases~\cite{NguyenQuantumSimulationCircular2018}.  The $\delta E_0$ and  $\delta \zeta$ energy shifts values in $H_S$ are listed in the first two lines of Table~\ref{tab:vdw}. All the parameters in the Hamiltonian have a $1/d^6$ distance dependence. Note that the full \XXZ\ Hamiltonian of Ref.~\cite{NguyenQuantumSimulationCircular2018} is obtained by adding a coherent microwave drive nearly resonant on the two-photon $\ket g\rightarrow\ket e$ transition. It amounts to a position-independent modification of $\delta \zeta$, that does not modify the spin-exchange dynamics considered in this work, and to the introduction of an effective transverse magnetic field Hamiltonian [$\propto \Omega (\hsigma_1^X+\hsigma_2^X)$]. We will not consider this more general situation here, namely we only consider the $\Omega=0$ situation. Note that the value of $\delta \zeta$ is given in the rotating frame where $\delta \zeta=0$ at infinite interatomic distance.

The spin Hamiltonian $\hat{H}_S$ splits the four-dimension Hilbert space of the two spins into three uncoupled blocks. Two correspond to the states $\ket{g,g}$ and $\ket{e,e}$. These states, besides energy shifts, have no spin dynamics and we can thus focus on the spin exchange evolution inside the last block, made up of the subspace $\{\ket{e,g},\ket{g,e}\}$, in which $\hat{H}_S$ reduces to a $2\times2$ matrix, denoted $\Htilde_S$. In this two-level subspace, we introduce a new set of Pauli matrices, $\hSigma_{X,Y,Z}$, with $\hSigma_Z = \ketbra{e,g}{e,g} - \ketbra{g,e}{g,e}$. Using the identities
\begin{align}
\hsigma_1^X \hsigma_2^X + \hsigma_1^Y \hsigma_2^Y & = 2[ \hsigma_1^+ \hsigma_2^- + \hsigma_1^- \hsigma_2^+] = 2\hSigma_X \nonumber \\
\hsigma_1^Z \hsigma_2^Z &= -1\,\quad \hsigma_1^Z + \hsigma_2^Z = 0\ ,
\end{align}
the reduced Hamiltonian reads:
\begin{align}
\frac{\Htilde_S}{\hbar} &= 2J \hSigma_X  - \Delta_0 \quad\text{with}\quad
\Delta_0 = J_z - \delta E_0\;.
\label{eq:HS2}
\end{align}
This Hamiltonian describes a spin-exchange Rabi oscillation at a frequency $4J$ between $\ket{e,g}$ and $\ket{g,e}$. The diagonal term $\Delta_0$ plays no role in this exchange dynamics and could a priori be removed. However, we must keep it here since it depends on the interatomic distance and, thus, participates to the spin-motion coupling.

As shown later, a strong entanglement between motion and spin requires an exchange interaction much larger than that reached in the $\Delta n=2$ situation described above. We thus consider also the case in which the spin model is encoded on two adjacent circular states, $\{\ket{g}=\ket{nC},\ket{e}=\ket{(n+1)C}\}$. In this $\Delta n=1$ situation, the spin exchange is a resonant, first order process in the dipole-dipole interaction. The Hamiltonian $\Htilde_S$ retains the same form, but $J$ overwhelms the position-dependent part of the second-order van der Waals interaction term, $\Delta_0$. The actual parameters values for $d=\SI{8}{\micro\meter}$ are given in the last line (\textbf{CATS}) of Table~\ref{tab:vdw}. In the following, we use the generalized form
\begin{equation}
 \Htilde_{S,m}=2J \hSigma_X  - \Delta_m, \quad \mathrm{with} \quad \Delta_m =  \delta_{m,3}\, \Delta_0,
\label{eq:reducedH}
\end{equation}
where $\delta$ is the Kronecker symbol, where all the coefficients are evaluated with an interatomic distance $d$, and where $m$ is the decay exponent of the interaction, that scales as $1/d^m$.

\subsection{Spin-motion Hamiltonian}
We now consider the coupling between spin and motion induced by the interatomic distance dependence of the spin-interaction terms. The positions of the atoms with respect to the center of each trap are represented by the operators  $\hat{x}_{1,2}$. We will write the dynamics in terms of the relative position operator $\hat{x} = \hat{x}_2-\hat{x}_1$, which vanishes when the two atoms are both at the trap centers (interatomic distance $d$). 

The spin Hamiltonian can then be written as
\begin{align}
V_m(\hat{x}) = U_m(\hat{x})
 \otimes \Htilde_{S,m}
\; \text{ with }\; U_m(\hat{x}) = \left(1+ \frac{\hat{x}}{d}\right)^{-m}\, ,
\label{eq:VvdW}
\end{align}
where $m$ is the decay exponent of the interaction. In the $\Delta n=1$ situation, $m=3$. In the $\Delta n=2$ situation, $m=6$. In section~\ref{sec:cats}, we will focus on the $m=3$ case. In all other sections we will consider the $m=6$ case.

 For small enough motions, the adjacent longitudinal trap wells in which the atoms are trapped can be considered as harmonic, with an oscillation frequency $\omega$. The interaction-free motion Hamiltonian of the two particles $i=1,2$ is then
\begin{equation}
\label{eq:harmonictrapping}
\hat{H}_i = \frac{\hat{p}_i^2}{2M} + \frac 12 M \omega^2 \hat{x}_i^2\ ,
\end{equation}
where $M$ is the atomic mass. 
Changing to the relative motion coordinates and dropping the center-of-mass part of the Hamiltonian, which decouples from the relative-motion part, one gets
\begin{align}
\hat{H} =  \frac{\hat{p}^2}{2\mu} + \frac 1 2 \mu \omega^2 \hat{x}^2 + U_m(\hat{x})\otimes\hbar\big( 2J \hSigma_X - \Delta_m \big)\ ,
\label{eq:Ham}
\end{align}
where $\mu = M/2$ is the reduced mass.
For later calculations, a convenient basis is that of the harmonic oscillator associated to the relative motion, with a natural length scale
\begin{equation}
x_0 = \sqrt{\frac{\hbar}{2\mu\omega}} \ .
\end{equation}
With the quantization $\hat{x} = x_0 (\ha^\dag+\ha)$ and $\hat{p} = (i\hbar/2x_0)(\ha^\dag - \ha)$, followed by the introduction of a dimensionless relative position
\begin{equation}
\hat{\xi} =  \hat{x}/x_0
\end{equation}
associated to the momentum $\hat{\pi} =  i(\ha^{\dag}-\ha)$, with $\comm*{\hat{\xi}}{\hat{\pi}} = 2i$,
 the Hamiltonian finally reads:
\begin{align}
\frac{\hat{H}}{\hbar} &= \omega\Big(\hat{n}+\frac12\Big) + \big(1+g\hat{\xi}\,\big)^{-m}\hspace{-1mm}\otimes\big( 2J \hSigma_X - \Delta_m\big) \ .
\label{eq:ham}
\end{align}
The two most important timescales in this Hamiltonian are $2\pi/\omega$, the period of the mechanical oscillations of the atoms, and $1/4J$, the period of the spin-exchange Rabi oscillations. The coupling strength between the position and spin degrees of freedom is controlled by
\begin{align}
g = \frac{x_0}{d}\;,
\end{align}
which is assumed to be much smaller than one. We 
note that $\ev*{\hSigma_X}$ is a conserved quantity since $[\hat{H},\hSigma_X]=0$.

The $U_m(\hat{x})$ operator can be expanded as a power series in the coupling constant $g$ to order $K$:
\begin{equation}
U_m(\hat{x})=\big(1+g\hat{\xi}\,\big)^{-m} = \sum_{k=0}^{K} c_m(k)g^k \hat{\xi}\,^k\;,
\end{equation}
with $c_m(k) = (-m)(-m -1)\cdots(-m-k+1)/k!$.
We will use numerical matrix inversion to treat the exact $K=\infty$ case. We check that the main source of error in the numerics originates in the mandatory truncation $N_{\text{max}}$ of the harmonic oscillator Hilbert space. 

For analytical approaches, we will consider  the linearized  operator ($K=1$) as well as the effect of the first non-linear term ($K=2$). Up to second order, we can also write
\begin{equation}
U_m(\hat{\xi}) \simeq  1-\eta\hat{\xi} + \frac{f}{2}\eta^2\hat{\xi}^2\,,\label{eq:expansion2}
\end{equation}
where  $f=1+1/m$ and where we define the effective Lamb-Dicke parameter, $\eta$, as
\begin{align}
\eta = g m\ .
\end{align}

Note that, since the dipole-dipole interaction has a fast variation with the average interatomic distance $d$, it results in forces, that could overwhelm the trapping force if the traps are weak enough. The stability of the trapping would then be questionable. We investigate this effect in Appendix~\ref{app:stability}. We show that, for weak harmonic traps, the action of the dipole-dipole forces could indeed lead to instabilities. However, for the rather tight traps considered in the following, the stability is not an issue.

The  independent parameters in this situation are $d$, $n$, $\Delta n$ and $\omega$, from which all relevant quantities can be computed. For instance, the coupling constant between spin and motion is expressed as
\begin{align}
g = \sqrt{\frac{\hbar}{M}}\frac{1}{d\sqrt{\omega}}\;.
\end{align}

For the sake of definiteness, we will consider in the following only  three sets of parameters designed by acronyms describing their main application
\begin{itemize} 
\item \textbf{XXZ } for the best realization of the pure spin-exchange model, with minor influence of atomic motion,
\item \textbf{THM } for a maximum impact of atomic motional temperature on the spin model and application to atomic motion thermometry,
\item \textbf{CATS }  for motion-spin entanglement and the generation of cat states.
\end{itemize}
The precise values of the parameters are given in Table \ref{tab:vdw}. To these parameters we may add, when relevant, the initial temperature of the atomic motion in the traps.  We will use a typical temperature range of $2$--$\SI{20}{\micro\kelvin}$ ($\SI{1}{\micro\kelvin} \equiv \SI{20.84}{\kilo\hertz}$). It can be experimentally reached with a final optical molasses cooling stage before preparation of circular states, eventually followed by additional cooling through adiabatic lowering of the laser trapping 
potential~\cite{Mehaignerie2023}.

\section{A simple intuitive approach}

\subsection{Interfering paths}
\label{sec:interferences}

\begin{figure}[t]
\centering
\includegraphics[width=\columnwidth,clip]{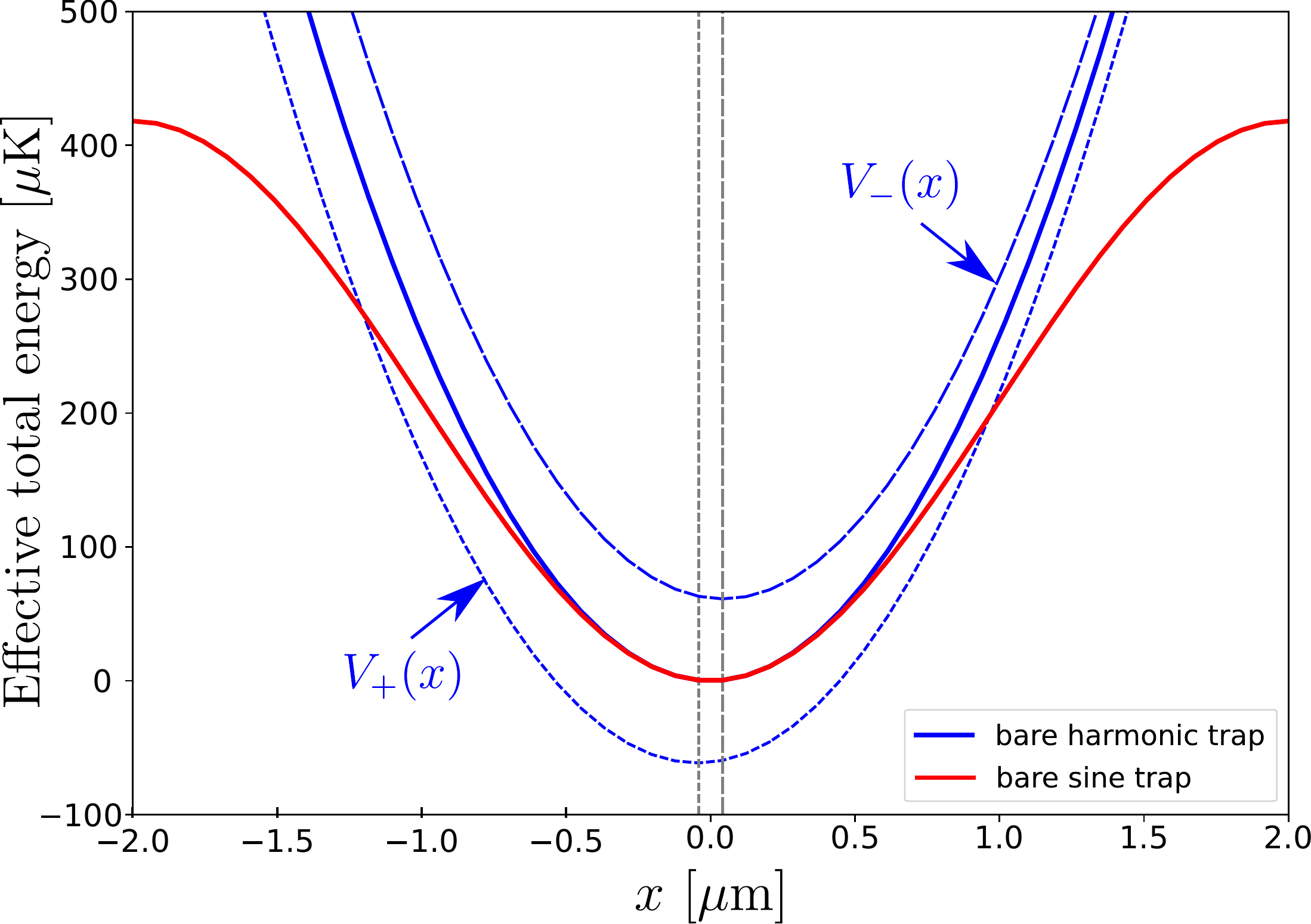}
\caption{Trapping potential and the corresponding effective potentials versus the  relative distance $x$ for the \bf CATS \rm set of parameters (see Table \ref{tab:vdw}). The solid red line is the actual sinusoidal ponderomotive potential, the solid blue line its harmonic approximation. The blue dotted and dashed lines are, respectively, the $V_{+}(x)$ and $V_{-}(x)$ effective potentials for each atomic state $\ket{\pm X}$. The black dashed and dotted thin vertical lines represent the centers of these effective potentials.}
\label{fig:exppotential}
\end{figure}

In this Section, we provide a first general view on the spin-motion dynamics, based on the quantum interference between two paths. 
Using the projectors on the two eigenstates of $\hSigma_X$, $\ket{\pm X} = (\ket{e,g}\pm\ket{g,e})/\sqrt{2}$  with eigenvalues $\pm 1$, we get
\begin{equation}
\hat{H} = \hat{H}_+\otimes\ketbra{X} 
+ \hat{H}_-\otimes\ketbra{-X}\ ,
\end{equation}
in which we recognize two different single-particle effective Hamiltonians for the relative position dynamics
\begin{equation}
\hat{H}_{\pm} =  \frac{\hat{p}^2}{2\mu} + V_{\pm}(\hat{x})\ ,
\end{equation}
with the two spin-state-dependent potentials
\begin{equation}
V_{\pm}(\hat{x}) \simeq \frac 12 \mu \omega^2 \hat{x}^2 + \hbar(\pm 2J - \Delta)U(\hat{x}) \;.\label{eq:Vpm}
\end{equation}
Note that, for the sake of simplicity, we have dropped the $m$ index for the potential $U_m$ and the energy $\Delta_m$, and replace them by $U$ and $\Delta$, respectively. 
The potentials $V_\pm$ are depicted in Figure \ref{fig:exppotential}. The trapping stability in presence of the dipole interaction forces, and hence the physical relevance of these shifted potentials, is discussed in Appendix~\ref{app:stability}. 

Let us assume that the initial state is the product state $\ket{\Psi(t=0)} = \ket{\psi_0(x)}\ket{e,g}$, where $\psi_0(x)$ is an arbitrary wavefunction. Due to the block diagonal form of the Hamiltonian, the evolving state simply reads
\begin{equation}
\ket{\Psi(t)} = \frac{1}{\sqrt{2}}\left(\ket{\psi_+(x,t)}\ket{+X} + \ket{\psi_-(x,t)}\ket{-X}\right)\ ,
\label{eq:psit}
\end{equation}
in which 
\begin{equation}
\ket{\psi_{\pm}(x,t)} = e^{-i\hat{H}_{\pm} t} \ket{\psi_0(x)}
\end{equation}
are the two wave packets, initially both equal to $ \ket{\psi_0(x)}$  but evolving under the two different $V_\pm$ potentials. 

Expressing the evolved state in the $\{\ket{e,g},\ket{g,e}\}$ basis, we find that $\ket{\psi_{\pm}(x,t)}$ involves two so-called ``cat states" for the relative position correlated to the two atomic states:
\begin{equation}
\ket{\Psi(t)} \propto
\frac{\ket{\psi_+}+\ket{\psi_-}}{\sqrt{2}}\ket{e,g} +
\frac{\ket{\psi_+}-\ket{\psi_-}}{\sqrt{2}}\ket{g,e} \ .
\label{eq:simplecat}
\end{equation}
Then, the probability $P_{eg}(t)$ that the atoms are back to the initial spin state $\ket{e,g}$ (the spin-exchange Rabi oscillation signal) is directly probing the interference between these two paths:
\begin{align}
P_{eg}(t) & = \frac{1}{2}
\Big[1 + \Re \braket{\psi_{-}(x,t)}{\psi_{+}(x,t)} \Big]
\label{eq:Peg} \\
&= \frac{1}{2}
\left[1 + \Re\left(\int \psi^*_{-}(x,t)\psi_{+}(x,t)\dd x  \right) \right]\;.
\end{align}
In the limit of very weak interactions, $g\to 0$, $U(\hat{x})$ reduces to the unity operator. The potentials $V_{\pm}(\hat{x}) $ have the same shape, but are shifted by  two different energies $\pm 2J -\Delta$. The wave packets $\ket{\psi_\pm}$ are identical, but evolve with different phase factors. We thus recover the ordinary spin-flip-flop Rabi oscillation:
\begin{equation}
P_{eg}(t) = \frac{1}{2}
\left[1 + \cos\left(4J t  \right) \right]\;.
\label{eq:standardrabi}
\end{equation}
One of the main goals of this paper is to understand how the coupling of the effective spin to the relative motion modifies this simple behavior.

\subsection{An oscillator coupled to a rotator}

Another simple way to understand the dynamics is to use Ehrenfest's theorem. 
The dynamical evolution of an observable expectation value, $A(t) = \expval*{\hat{A}}{\psi(t)}$, is given by $\dot{A} = (i/\hbar)\expval*{\comm*{\hat{H}}{\hat{A}}}$.
For the relative position, one gets
\begin{align}
\ddot{x} + \omega^2x &= - \frac{\hbar}{\mu}\left\langle{\dv{U}{\hat{x}}} (2J\hSigma_X-\Delta)\right\rangle,
\label{eq:xg}
\end{align}
i.e., an oscillator forced by its coupling to the effective spin expectation value with  components  $\vec{s} = \ev*{\vec{\Sigma}}$. Since $\dot{s}_X = 0$, the spin dynamics only takes place in a plane parallel to  the $YZ$ one.
We introduce a phase space operator
\begin{equation}
\hat{S} = \hSigma_Z + i\hSigma_Y\;,\;\text{with }\comm*{\hSigma_X}{\hat{S}} = -2\hat{S}\ ,
\end{equation}
which physically follows the evolution in the spin phase space.
The equation of motion for the overlap function $S(t) = \expval*{\hat{S}}=\braket{\psi_-}{\psi_+}$ [see \eqref{eq:psit}]   reads
\begin{align}
\dot{S}(t) &= -i4J \expval*{U(\hat{x})\hat{S}} \,.
\label{eq:Sdot}
\end{align}
In the following, we study the coupling between the oscillator $\hat{x}$ and the rotator $\hat{S}$ as one progressively expands  $U(\hat{\xi})$  to orders one (linear coupling) and two (first non linear term) [see \eqref{eq:expansion2}].

\section{The linear coupling limit}

\subsection{The potentials structure}

We start with the linearized interaction potential ($K=1$)
\begin{equation}
U(\hat{\xi}) \simeq 1-\eta \hat{\xi}\;.
\label{eq:UK1}
\end{equation}
Rewriting the two spin-dependent potentials as
\begin{equation}
\label{eq:Vpm_linear}
V_{\pm}(\hat{x}) = \frac 12 \mu \omega^2 (\hat{x}-x_0^{\pm})^2
 \pm \frac{\hbar \Omega}{2} - \hbar\Delta - \hbar \omega(\alpha^2 + \alpha_z^2)\;,
\end{equation}
we see that the potentials are shifted in position and energy with respect to each other (see Fig.~\ref{fig:exppotential}). Each potential is now centered around a spin-dependent minimum
\begin{equation}
\label{eq:x0pm}
x_0^{\pm} =2\, x_0\alpha^\pm= 2x_0(\pm \alpha - \alpha_z)\;,
\end{equation}
with the two dimensionless shifts in phase space
\begin{equation}
\alpha = \eta \frac{2J}{\omega}\;\text{and }\;
\alpha_z = \eta \frac{\Delta}{\omega}\;.
\end{equation}
The energy difference between the minima of the two potential wells, $V_{\pm}(x) $, corresponds to the Rabi frequency
\begin{equation}
\label{eq:Omega}
\Omega = 4J\left(1+2\eta^2\frac{\Delta}{\omega}\right)\;,
\end{equation}
 only modified to second order in $\eta$. 
The energy spectrum is straightforwardly obtained
from \eqref{eq:Vpm_linear}:
\begin{equation}
\label{eq:Ens}
\frac{E(n,s)}{\hbar} = \omega\left(n+\frac{1}{2}\right) + \frac{\Omega}{2}s  
-\Delta - \omega(\alpha^2+\alpha_z^2)\ ,
\end{equation}
with $s=\pm 1$ and $n\in\mathbb{N}$.
Since the potentials remain parabolic, the relative motion is still harmonic. From the Ehrenfest equation \eqref{eq:xg}, taking as initial conditions $s_X=\cos\theta_0$ and $\dot{x}=0$, we get
\begin{align}
\ddot{x} + \omega^2 (x-\bar{x}) &=0\ ,
\label{eq:xehrenfest}
\end{align}
with 
\begin{equation}
\bar{x} = 2x_0(\alpha \cos\theta_0 - \alpha_z)\ .
\end{equation}
In this limit, the dynamics of the position expectation value is independent of the spin dynamics. A simple physical interpretation of the Rabi frequency
formula \eqref{eq:Omega} is that it corresponds to making the simplest approximation $\hat{x} \simeq \bar{x}$ in \eqref{eq:Sdot} assuming \eqref{eq:UK1} and $\cos\theta_0 = 0$.

\subsection{Diagonalization using Polaron transformation}
\label{sec:decoupling}

Shifted harmonic potentials can be diagonalized with displacement operators. We need to condition them to the spin operator in order
to diagonalize the full linearized Hamiltonian
\begin{align}
\frac{\hat{H}}{\hbar} &= \omega\Big(\hat{n}+\frac12\Big) 
+ \big(1-\eta\hat{\xi}\big)\otimes\big( 2J \hSigma_X - \Delta\big)\;.
\label{eq:H1}
\end{align}
At the operator level, this is performed using a unitary polaron transformation 
\begin{equation}
\label{eq:Utransfo}
\hat{\cal U }(\alpha,\alpha_z) = \hat{D}(\alpha\hSigma_X-\alpha_z)\;,
\end{equation}
in which $\hat{D}$ is the Glauber's displacement operator~\cite{GlauberCoherentincoherentstates1963}. In the transformed Hamiltonian $\polaronH = \hat{\cal U }^\dagger(\alpha,\alpha_z) \hat{H} \hat{\cal{U} }(\alpha,\alpha_z)$, spin and positions are decoupled:
\begin{align}
\frac{\polaronH}{\hbar} &=\quad \omega \Big(\hat{n}+\frac12\Big) + \frac{\Omega}{2} \hSigma_X  
\label{eq:H1tilde}
 -\Delta - \omega(\alpha^2+\alpha_z^2)\;.
\end{align}
The energies of this Hamiltonian are given by \eqref{eq:Ens}.
Using this diagonal form, the time evolution operator in the original basis reads
\begin{align}
e^{-it\hat{H}} &= \hat{D}\left[(\alpha\hSigma_X-\alpha_z)(1-e^{-i\omega t})\right] 
e^{-i\omega t\hat{n}}  \nonumber\\
&\quad \otimes e^{-i(\alpha\hSigma_X-\alpha_z)^2\sin(\omega t)}e^{-i\frac{\Omega t}{2} \hSigma_X}\ .
\end{align}
A direct application of this result is to compute the spin dynamics when starting in a Fock state $\ket{n}$ for the relative position and in $\ket{e,g}$ for the atomic state. The phase space operator expectation value, $S(t)$, is then given by 
\begin{align}
S_n(t) = e^{-2\abs{\gamma(t)}^2} L_n(\abs{2\gamma(t)}^2) e^{i(\Omega t - \Theta(t))}\ ,
\label{eq:fock}
\end{align}
in which $\gamma(t) = \alpha(1-e^{-i\omega t})$, 
$\Theta(t) = 4\alpha\alpha_z\sin(\omega t)$ and $L_n$ is the $n^{\text{th}}$ Laguerre polynomial.
The Rabi oscillation contrast is thus governed by the oscillating
$\gamma(t)$ function.

\subsection{Cat states and Rabi oscillations}

\begin{figure}
\includegraphics[width=0.9\columnwidth]{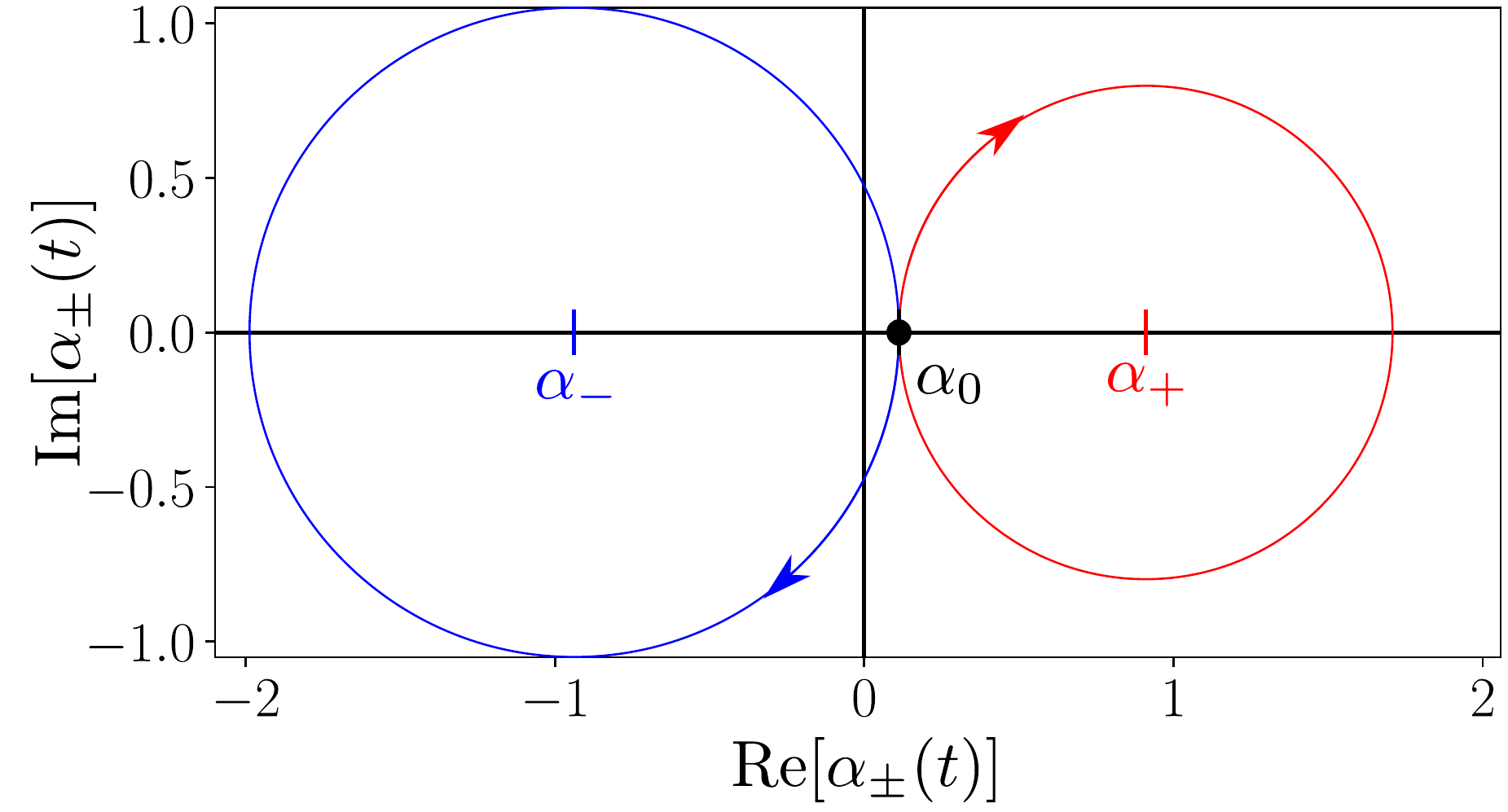}
\caption{Phase space trajectories of the states $\ket{\psi_\pm(t)}$ corresponding to an initial coherent state $\ket{\alpha_0}$, in the linear coupling ($K=1$) case. The values of $\alpha^{\pm}$ correspond to the \textbf{CATS} parameter set of Tab.~\ref{tab:vdw}, and, here, $\alpha_0 = 1+\alpha_-$.}
\label{fig:circles}
\end{figure}

The linear coupling limit allows us to compute explicitly the cat state
\eqref{eq:psit}. For the sake of definiteness, we consider the initial state $\ket{\psi_0(x)}$ to be
 a coherent state $\ket{\alpha_0}$ for the position so that
$\ket{\Psi(t=0)} = \ket{\alpha_0}\ket{e,g}$. 
The $\ket{\psi_\pm(x,t)}$ wavefunctions are then coherent states, too. Their amplitudes evolve, in the oscillator phase space, on circles centered around $\alpha^{\pm}$ and of radius $|\alpha_0-\alpha^{\pm}|$ (see Fig.~\ref{fig:circles}). More formally, we get
\begin{align}
\ket{\psi_{\pm}(x,t)} &= e^{\mp i\frac{\Omega}{2}t + i\theta_\pm(t)}\ket{\alpha^{\pm}+(\alpha_0-\alpha^{\pm})e^{-i\omega t}} \\
& =  e^{\mp i\frac{\Omega}{2}t + i\theta_\pm(t)}\ket{\gamma_0(t) \pm \gamma(t)}\ ,
\end{align}
in which we have defined
\begin{align}
\label{eq:gamma0}
\gamma_0(t) &= \alpha_0 e^{-i\omega t}-\alpha_z(1-e^{-i\omega t})\;,\\
\label{eq:gammar}
\gamma(t) &= \alpha(1-e^{-i\omega t})\;.
\end{align}
The phases $\theta_\pm(t)$, resulting from the basis change used in the polaron transformation, read, assuming $\alpha_0 \in \mathbb{R}$,
\begin{equation}
\theta_\pm(t)=  \left(\alpha^\pm \alpha_0 - {\alpha^\pm}^2\right) \sin \omega t \ .
\end{equation}
Using \eqref{eq:Peg}, the spin flip-flop Rabi oscillation can be explicitly computed. We get
\begin{equation}
\label{eq:Pesimple}
P_{eg}(t) = \frac{1}{2}
\left\{1 + e^{-4\alpha^2(1-\cos\omega t)} \cos\left[\Omega t -\Theta(t)\right] \right\}\;,
\end{equation}
with
\begin{equation}
\label{eq:Theta}
\Theta(t) = 4\alpha(\alpha_z+\alpha_0)\sin(\omega t)\;.
\end{equation}
Compared to \eqref{eq:standardrabi} describing the motion-free spin exchange, the frequency of the Rabi oscillation  is slightly altered as  shown in \eqref{eq:Omega}. The oscillation is also phase-shifted by the time-dependent phase $\Theta(t) $ oscillating at the trap oscillation frequency. Finally, the most important modification due to the spin-motion coupling is the periodic contrast function $ \exp [-4\alpha^2(1-\cos\omega t)]$, which oscillates at the trap frequency, too. 
For a small displacement $\alpha$ (i.e. for a small spin-motion coupling $\eta$), this function is close to 1. In the limit of vanishing coupling, we thus recover the simple flip-flop Rabi oscillation. On the contrary, for $|\alpha|>1$, this function nearly collapses periodically. The vanishing Rabi oscillation contrast reveals a full entanglement of the spin and motional degrees of freedom, in a Schr\"odinger-cat state involving two well-separated coherent components in the oscillator phase space. We now explore numerically this phenomenon in realistic conditions.

\subsection{Numerical results in experimentally realistic conditions}
\label{sec:cats}

What are the experimental conditions to observe such a cat state? First of all, we need a large enough $\alpha$ to separate the two coherent components in the oscillator phase space, but we also have to remain in the weak coupling regime $g \ll 1$ so that the linear dependence of the coupling with interatomic distance is valid. With the spin coded on a two-photon transition between two Rydberg states $n$ and $n+2$, we find that it is difficult to separate the two cat components without getting an annoying contribution of the non-linear effects at large $\alpha$. 

\begin{figure*}[t]%
\centering
\includegraphics[width=0.82\textwidth,clip]{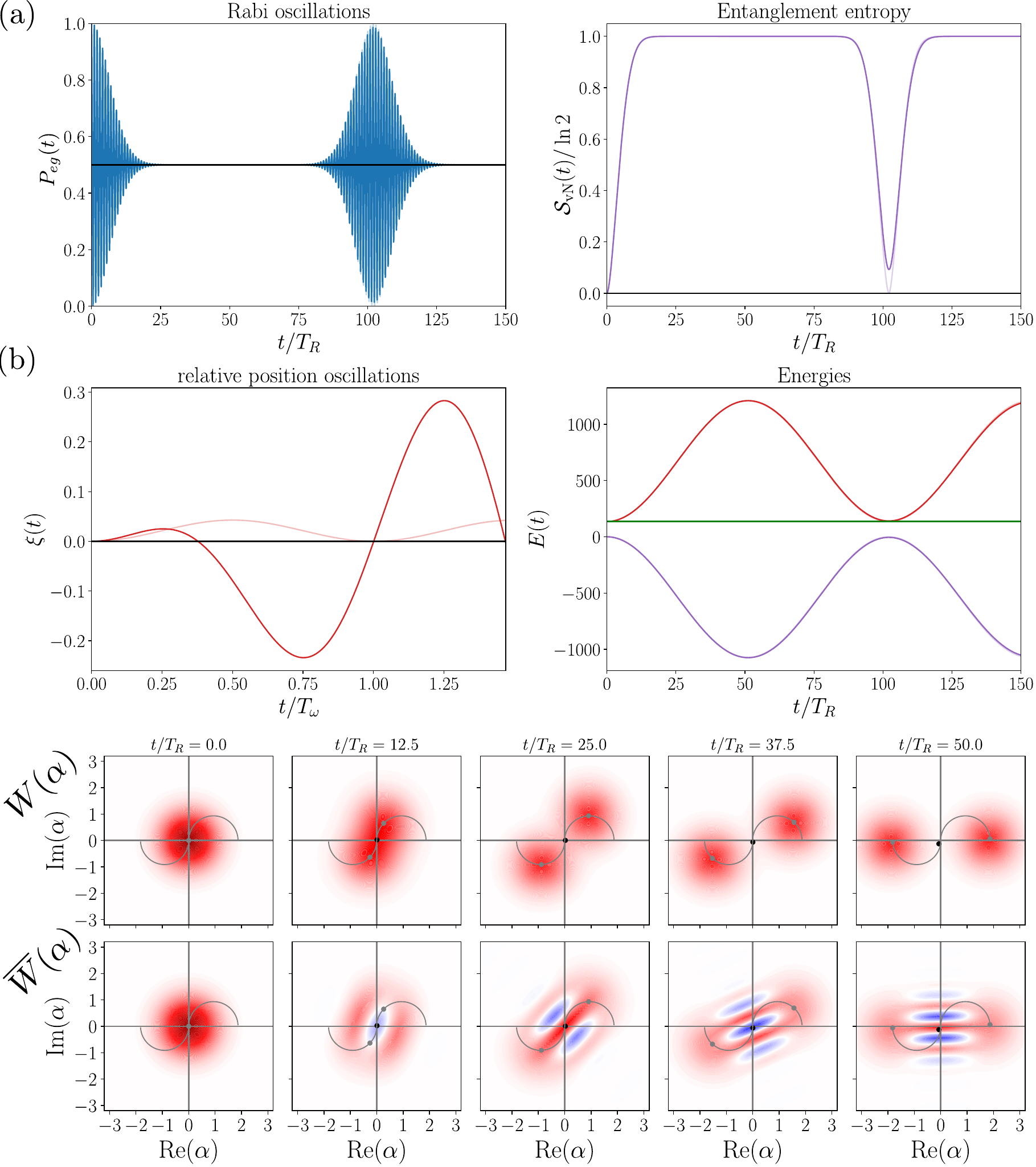}
\caption{Example of cat states appearing in the joint spin-motion dynamics. The parameters are those of the \textbf{CATS} line in Table~\ref{tab:vdw}. The initial atomic temperature in the traps is $T=\SI{2}{\micro\kelvin}$. These simulations are performed including the exact, non-linear dependence of the van der Waals interactions with distance ($K=\infty$). Curves with lighter colors are for $K=1$ (linear coupling) with the same parameters. We use blue for quantities relative to the effective spin, red for those related to the relative position, and violet for mixed spin-position quantities. (a) Top row: spin-exchange Rabi oscillation (probability to find the atoms in the state $\ket{e,g}$) and spin-motion entanglement entropy. Second row: relative position and spin/motion energies (see text) versus time. The green line is the total energy.
(b) First row: snapshots of the motional Wigner function $W(\alpha)$ during the first spin-exchange Rabi oscillation period. Second row: projected Wigner function
$\overline{W}(\alpha)$  at the same times. In these two rows,  the green lines represent the two circular trajectories $\xi_{\pm}(t)$ of the cat state's coherent components amplitudes (see Fig. \ref{fig:circles}).}%
\label{fig:cat}
\end{figure*}

Consequently, we rather consider a $\Delta n=1$ situation (\bf CATS \rm  parameters in Tab.~\ref{tab:vdw}). As stated above, the interatomic dependence of the spin exchange term in the Hamiltonian corresponds to $m=3$, while all other van der Waals terms have a $m=6$ dependence. The analytic solutions above thus only apply when using the expression of $\Delta$ defined in~\eqref{eq:reducedH}, namely when canceling the $\Delta$ terms. For the numerical approach, instead, we use the full position dependence of the spin-motion Hamiltonian:
\begin{align}
V_{\mathrm{full}}(\hat{x})/\hbar &= 
U_3(\hat{x}) \otimes  2J \hSigma_X - U_6(\hat{x})\otimes  \Delta_0\ ,
\label{eq:VvdW_m3}
\end{align}
and we take into account the exact form of the potentials ($K=\infty$). The initial motional state is a thermal equilibrium (centered on $\alpha_0=0$) at a temperature $T=\SI{2}{\micro\kelvin}$. From the simple theoretical model, we predict $\alpha_+=0.91$ and $\alpha_-=-0.94$. The Hilbert space for motion is truncated to a maximum of 100 phonon states. We have checked, that the numerical precision is not limited by this truncation.

The results of this simulation are plotted in Fig.~\ref{fig:cat}.
For the sake of clarity, we use the color blue for quantities relative to the effective spin, red for those related to the relative position, and violet for mixed spin-position quantities. 
The figure shows the main relevant observables: the Rabi oscillation signal, $P_{eg}(t)$, and the entanglement entropy between the oscillator and the rotator, evaluated using the von Neumann entanglement entropy
\begin{equation}
S_{\rm vN} = -\Tr\left[\hat{\rho}_S(t)\ln\hat{\rho}_S(t)\right]\ ,
\label{eq:Svn}
\end{equation}
where $\hat{\rho}_S(t)$ is the effective-spin density matrix.
The maximal value of $S_{\rm vN}$ is $\ln 2$. We also plot $\xi(t)$, the dimensionless relative position. Regarding energy exchanges, we split
the Hamiltonian into three energy terms $\hat{H} =\hbar( \hat{H}_o + \hat{H}_S + \hat{H}_{oS})$ in which 
\begin{align}
\hat{H}_o &= \omega \hat{n}\ ;\\ \hat{H}_S &= 2J \hSigma_X - \Delta_0\ ;
\label{eq:energies1} \\
\hat{H}_{oS} &= [U_3(\hat{\xi})-1)]\otimes 2J \hSigma_X
-[U_6(\hat{\xi})-1]\otimes\Delta_0\ .
\label{eq:energies2}
\end{align}
They are respectively the relative-motion energy, the spin energy and the spin-position coupling energy, which vanishes when $g\to 0$.
While the total energy (in green) and the spin energy are always conserved, there is some energy exchange between the relative motion (in red) and the coupling energy (in violet).
For this set of parameters, the Rabi period $T_R=2\pi/\Omega$ is much shorter than the position oscillation period $T_\omega=2\pi/\omega$. The Rabi oscillations and entanglement entropy nicely display a collapse and revival phenomenon with, in between, a plateau signaling the cat formation.

Beyond these mean-values and in order to probe the cat formation, the motion in phase space is presented through two Wigner functions. First, the standard Wigner function defined using the number parity operator $\hat{\Pi} = (-1)^{\hat{n}}$ as
\begin{align}
W(\alpha) &= \frac{1}{\pi}\Tr{\hat{\rho}_o \hat{D}(\alpha) \hat{\Pi} \hat{D}^\dag(\alpha) }  \ ,
\end{align}
with $\alpha = (x + ip)/\sqrt{2}$, $\hat{D}(\alpha)$ the displacement operator and $\hat{\rho}_o=\Tr_S \hat{\rho}$ the reduced density matrix for the position. Second, we see in \eqref{eq:simplecat} that the cat coherence should appear in the projected Wigner function 
\begin{align}
\overline{W}(\alpha) &= \frac{1}{\pi}\Tr{\bra{e,g}\hat{\rho}\ket{e,g} \hat{D}(\alpha) \hat{\Pi} \hat{D}^\dag(\alpha) }  \ ,
\end{align}
in which the oscillator density matrix is conditioned on the atomic state $\ket{e,g}$. Both Wigner functions are plotted in Fig.~\ref{fig:cat} for various times. For $36 \lesssim t/T_R \lesssim 60$, half-way before the Rabi revival, we clearly observe a well formed cat state, signaling the entanglement between the oscillator and the rotator. The size of such a cat is usually measured by the square of the distance between its components in phase space. We get here a maximum value of $13.6$ phonons, corresponding to a rather large animal.

\section{Finite temperature}

In this Section, we extend the analytical solution of the linear coupling model ($K=1$) to a finite initial atomic motion temperature. We also assume a more general initial spin state. We compute analytically all relevant observables and get direct estimates for Rabi contrast as a function of realistic experimental parameters. We finally show that a careful choice of these parameters turns the spin exchange into a precise thermometer for the initial atomic motional state, with interesting potential applications.

\subsection{Initial state}

We consider that the spin degree of freedom is prepared in a pure state.  Using a Bloch sphere representation of the spin $\Sigma$ with quantization axis along $X$ (note the unusual choice of orientation), we define the angles $(\theta_0,\phi_0)$ so that the initial spin state is
\begin{align}
\ket{\varphi}& = c_+ \ket{+X} + c_-\ket{-X}\ ,
\end{align}
with
\begin{align}
 c_+ = \cos\left(\frac{\theta_0}{2}\right)
\quad\text{and}\quad c_- = \sin\left(\frac{\theta_0}{2}\right) e^{i\phi_0}\;.
\end{align}
In density matrix form,  $\hat{\rho}_S = \ketbra{\varphi}$. The initial state $\ket{e,g}$, used in previous sections, is recovered when $\theta_0=\pi/2$ and $\phi_0=0$.

For the relative position, we consider a displaced thermal state.  At a finite temperature $T$, the phonon number in the harmonic well is $\bar{n} = 1/(e^{T_h/T}-1)$ with $T_h = \hbar \omega / k_B$ being the natural temperature scale (its relevant values are given in table \ref{tab:vdw}). The initial thermal density matrix is then
\begin{equation}
\hat{\rho}_\beta 
	= \frac{1}{1+\bar{n}}  \sum_{n=0}^\infty \left(\frac{\bar{n}}{1+\bar{n}}\right)^n \dyad{n} \;.
\end{equation}
We consider in addition an initial coherent displacement by an amplitude $\alpha_0$, as before. The displaced thermal state density matrix is $\hat{\rho}_\beta(\alpha_0) = \hat{D}(\alpha_0)\hat{\rho}_\beta \hat{D}^\dag(\alpha_0)$. Eventually, the full initial state is the factorized state 
\begin{equation}
\hat{\rho}(t=0) = \hat{\rho}_S \otimes \hat{\rho}_{\beta}(\alpha_0)\;.
\end{equation}

\subsection{Observables}

During the time evolution, we extract analytically and numerically all relevant quantities from the density matrix $\hat{\rho}(t)$. 
For the effective spin, $\vec{\Sigma}$, we compute all components of $\vec{s} = \ev*{\vec{\hSigma}}$, or, equivalently, the reduced spin density matrix $\hat{\rho}_S(t) = \Tr_{o} \hat{\rho}(t)$, where $\Tr_{o} $ denotes the trace over the mechanical oscillator degree of freedom. 
We already noticed that $s_X$ is a conserved quantity since $[H,\hSigma_X]=0$. The probability $P_{eg}(t) $ that the first atom is in an excited state (the Rabi oscillation signal) is given by
\begin{equation}
P_{eg}(t) = \Tr(\hat{\rho}(t)\ketbra{e,g}) = \frac12\left[1 + s_Z(t)\right]\;.
\end{equation}
For the relative position, we compute the averages
\begin{align}
\xi(t) &= \ev*{\ha^{\dag}+\ha} \;,\quad
\pi(t) = \ev*{i(\ha^{\dag}-\ha)} 
\end{align}
as well as the mean occupation of the oscillator $n(t) = \ev{\hat{n}}$. Finally, we also compute the entanglement entropy defined in Eq.~\eqref{eq:Svn}.

\subsection{Exact solution in linear coupling}

We can use the results of the polaron transformation in Sec.~\ref{sec:decoupling}, after which $\polaronH$ gives a trivial dynamics since spin and position are decoupled. Formally, the calculation boils down to a basis change 
\begin{equation}
\hat{\rho}(t) = e^{-i\hat{H}t}\hat{\rho}(0)e^{i\hat{H}t}
= \hat{\cal{U} } e^{-i\polaronH t} \hat{\cal U }^{\dagger}\hat{\rho}(0) \hat{\cal{U} } e^{i\polaronH t} \hat{\cal{U} }^{\dagger}\ ,
\end{equation}
that eventually entangles the spin and position degrees of freedom. 
The resulting state reads
\begin{align}
\hat{\rho}(t) &= \quad \abs{c_+}^2  \hat{D}(\gamma_0+\gamma)\hat{\rho}_\beta \hat{D}^{\dag}(\gamma_0+\gamma)\ketbra{X} \label{eq:rhooft} \\
			 &+ \abs{c_-}^2 \hat{D}(\gamma_0-\gamma)\hat{\rho}_\beta \hat{D}^\dag(\gamma_0-\gamma)\ketbra{-X} \nonumber \\
			 &+ c_+^*c_- e^{i\Omega t-i\vartheta(t)}
\hat{D}(\gamma_0-\gamma)\hat{\rho}_\beta \hat{D}^\dag(\gamma_0+\gamma)			 
			  \ketbra{-X}{X} \nonumber \\
			  \nonumber
			 &+ c_-^*c_+ e^{-i\Omega t+i\vartheta(t)}
\hat{D}(\gamma_0+\gamma)\hat{\rho}_\beta \hat{D}^\dag(\gamma_0-\gamma)	\ketbra{X}{-X}\ ,
\end{align}
in which one recovers the Rabi frequency $\Omega$. The density matrix coefficients display the overlap and entanglement between two coherent states, with amplitudes evolving on the circles $\gamma_0(t)\pm\gamma(t)$ in phase space [$\gamma_0(t)$ and $\gamma(t)$ are given by Eqs.~\eqref{eq:gamma0} and \eqref{eq:gammar}].
The relative phase shift $\vartheta(t) $ in \eqref{eq:rhooft} reads
\begin{equation}
\vartheta(t) = 2\alpha(2\alpha_z+\alpha_0)\sin(\omega t)\ .
\end{equation}
The spin dynamics is obtained by tracing out the relative position:
\begin{align}
\hat{\rho}_S(t) =& \quad \abs{c_+}^2 \ketbra{X} + \abs{c_-}^2 \ketbra{-X} \nonumber \\
			 &+ C(t)\left[c_+^*c_- e^{i\Omega t-i\Theta(t)}
			  \ketbra{-X}{X}\right. \nonumber\\
			 &\qquad\quad\left.+ c_-^*c_+ e^{-i\Omega t+i\Theta(t)}\ketbra{X}{-X} \right]\;,
\end{align}
in which we have introduced the contrast  function
\begin{align}
C(t) &= e^{-2(2\bar{n}+1)\abs{\gamma(t)}^2}
= e^{-4(2\bar{n}+1)\alpha^2(1-\cos\omega t)}\ ,
\label{eq:Ct}
\end{align}
and where the phase $\Theta(t)$ is defined in \eqref{eq:Theta}. 
From this density matrix, we compute the average spin components, which evolve as
\begin{align}
s_X(t) &= \cos\theta_0\\ 
s_Y(t) &= C(t)\sin\theta_0\sin[\Omega t- \Theta(t) +\phi_0]\\
s_Z(t) &= C(t)\sin\theta_0\cos[\Omega t- \Theta(t) +\phi_0]\ .
\end{align}
We obtain finally the extension of the spin-exchange Rabi oscillation signal [Eq.~\eqref{eq:Pesimple}] to a finite initial temperature 
\begin{equation}
P_{eg}(t) = \frac{1}{2}\left\{{1 + C(t)\;\sin\theta_0\cos\left[\Omega t  +\phi_0-\Theta(t)\right]}\right\}\ ,
\label{eq:Pet}
\end{equation}
where the Rabi oscillations are  modulated by the contrast $C(t)$. 
In addition, we also derive explicitly the entanglement entropy between the spin and position degrees of freedom:
\begin{align}
\mathcal{S}_{\text{vN}}(t) = -\lambda_+(t)\ln\lambda_+(t) -\lambda_-(t)\ln\lambda_-(t)\ ,
\end{align}
where the eigenvalues $\lambda_\pm$ of the reduced density matrix $\hat{\rho}_S(t)$ read
\begin{align}
\lambda_{\pm}(t) = \frac{1}{2}\left[1\pm\sqrt{\cos^2\theta_0 + \sin^2\theta_0\,C^2(t)} \right]\ .
\end{align}
A measurement of the contrast function $C(t)$ thus gives an indirect probe of the entanglement in the system. Maximum entanglement $\mathcal{S}_{\text{vN}} = \ln 2$ is reached when $\cos\theta_0=0$ and $C(t) \to 0$, which occurs in the linear limit when the coupling displacement $\alpha$ gets sufficiently large, as is the case for the \textbf{CATS} parameters.

\begin{figure}[t]%
\includegraphics[width=0.95\linewidth]{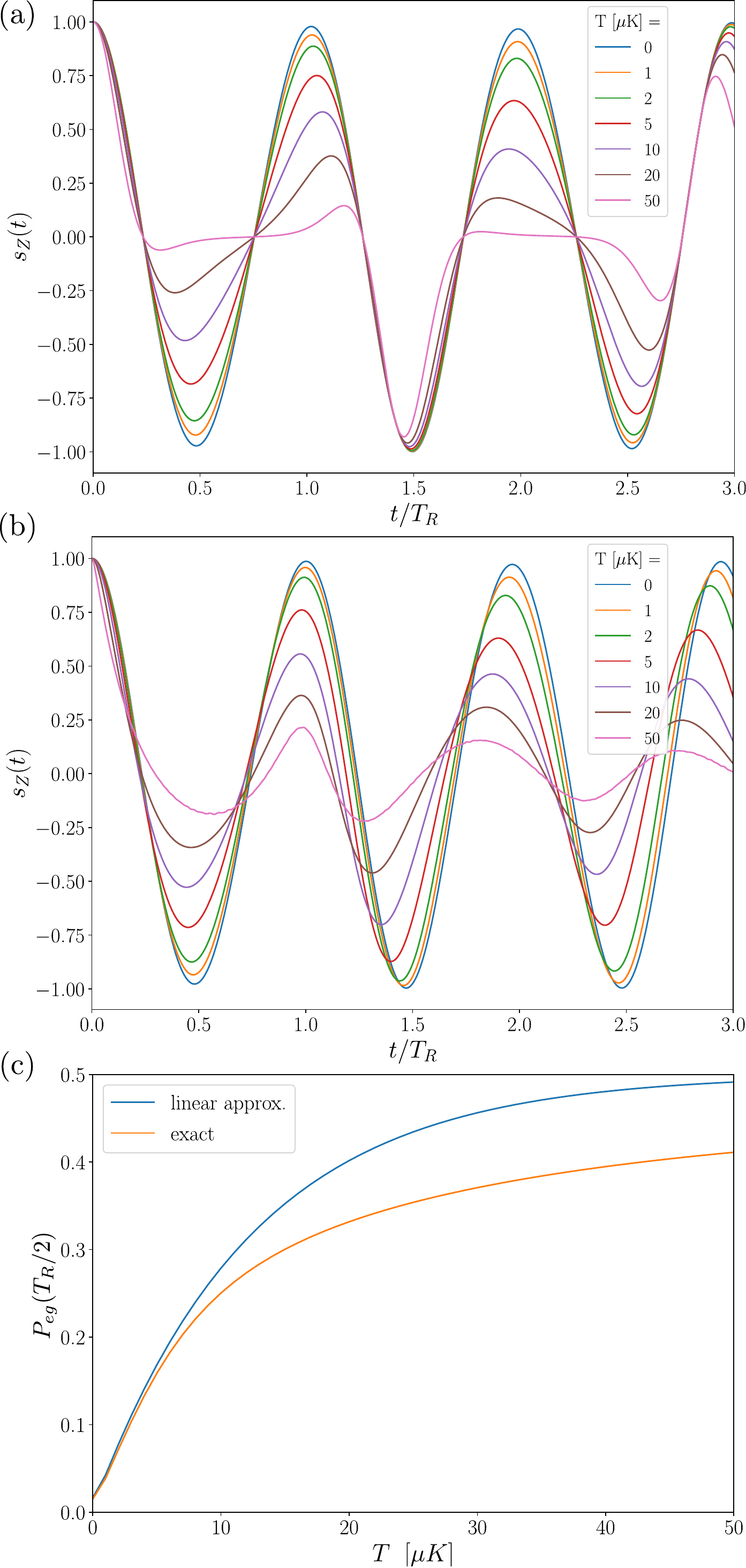}
\caption{Atomic motion thermometry with spin exchange Rabi oscillations. The experimental parameters are those of line \textbf{THM} from Table~\ref{tab:vdw}. 
{(a)} Analytical result for the linear coupling. 
{(b)} Exact numerical result including non-linear effects.
{(c)} Probability $P_{eg}(T_R/2)$ at the first minimum of $P_{eg}(t)$ used as a thermometer.}%
\label{fig:thermometry}%
\end{figure}

As detailed in Appendix~\ref{app:formulas}, we can also compute explicitly
all observables related to the relative position and energy.
Physically, observables involving $\hSigma_X$ result in a combination of the two trajectories contributions, while observables involving $\hSigma_Y$ and $\hSigma_Z$ capture interferences between the two paths and involve the contrast function $C(t)$. This naturally generalizes to the finite temperature case the considerations of Sec.~\ref{sec:interferences}. As an example, one has
\begin{align*}
\xi(t) =& \; 2\Re\left[\abs{c_+}^2\left(\gamma_0(t)+\gamma(t)\right) + \abs{c_-}^2(\gamma_0(t)-\gamma(t))\right] \nonumber\\
\expval*{\hat{\xi}\hat{S}}(t) = & \;
4c_-c_+^*C(t)\; e^{i[\Omega t-\Theta(t)]} \\
&\qquad \times \left[\Re(\gamma_0+\gamma)-\gamma(\bar{n}+1)+\gamma^*\bar{n}\right]\;.
\end{align*}
More explicitly, the mean relative position reads
\begin{align}
\xi(t) &= \bar{\xi} + (\xi_0 - \bar{\xi})\cos(\omega t)\ ,
\end{align}
with $\bar{\xi} = 2(\alpha\cos\theta_0 - \alpha_z)$ and
$\xi_0 = 2\alpha_0$.
This oscillating behavior with a constant shift $\bar{\xi}$ shows, that
the previous result \eqref{eq:xehrenfest} extends to finite temperature and is 
thus temperature-independent in the linear coupling limit.

\subsection{Application to atomic motion temperature measurement}

The above results suggest, that the spin-motion observables could be used for a precise determination of the initial temperature $T$ or, equivalently, of the mean phonon number $\bar{n}$ for the atoms in their traps. Indeed, the contrast function $C(t)$ and thus, for instance, the Rabi signal $P_{eg}(t)$, are dependent on  $\bar{n}$. This dependence is minimized for a large $\omega/J$ ratio, an optimal condition for the observation of nearly unperturbed spin exchange, as shown in the next section. For a weak trap (compatible with the stability of the system), i.e., for a small $\omega/J$ ratio, the temperature dependence is much higher and the spin exchange signal becomes a good thermometer for the initial atomic temperature. 

In order to exploit this sensitivity, while remaining in the realm of the linear coupling, we propose  to use the set of parameters \bf THM \rm of Tab.~\ref{tab:vdw} and focus on the first oscillations. The spin exchange oscillation signals are displayed in Fig.~\ref{fig:thermometry} for initial temperatures ranging from zero to $\SI{50}{\micro\kelvin}$. Figure~\ref{fig:thermometry}(a) presents the analytical solution in the linear regime and Fig.~\ref{fig:thermometry}(b) an exact numerical simulation taking into account the exact distance dependence ($K=\infty$). Both figures show a clear dependence of the $P_{eg}(t)$ signal with temperature, particularly in the region of the first minimum. Their comparison shows that, for such a weak $\omega$ value, the nonlinear effects already play an important role.

Figure~\ref{fig:thermometry}(c)  shows the first minimum of the Rabi oscillation signal, $P_{eg}^*=P_{eg}(T_R/2)$, as a function of temperature. The linear and exact results are in good agreement for low enough temperatures, with a common slope of about $\SI{2.2}{\%/\micro\kelvin}$. A determination of $P_{eg}^*$ at the \% uncertainty level thus provides an interesting $1$-$\si{\micro\kelvin}$ precision on the temperature. 

\begin{figure}[t]%
\includegraphics[width=\linewidth]{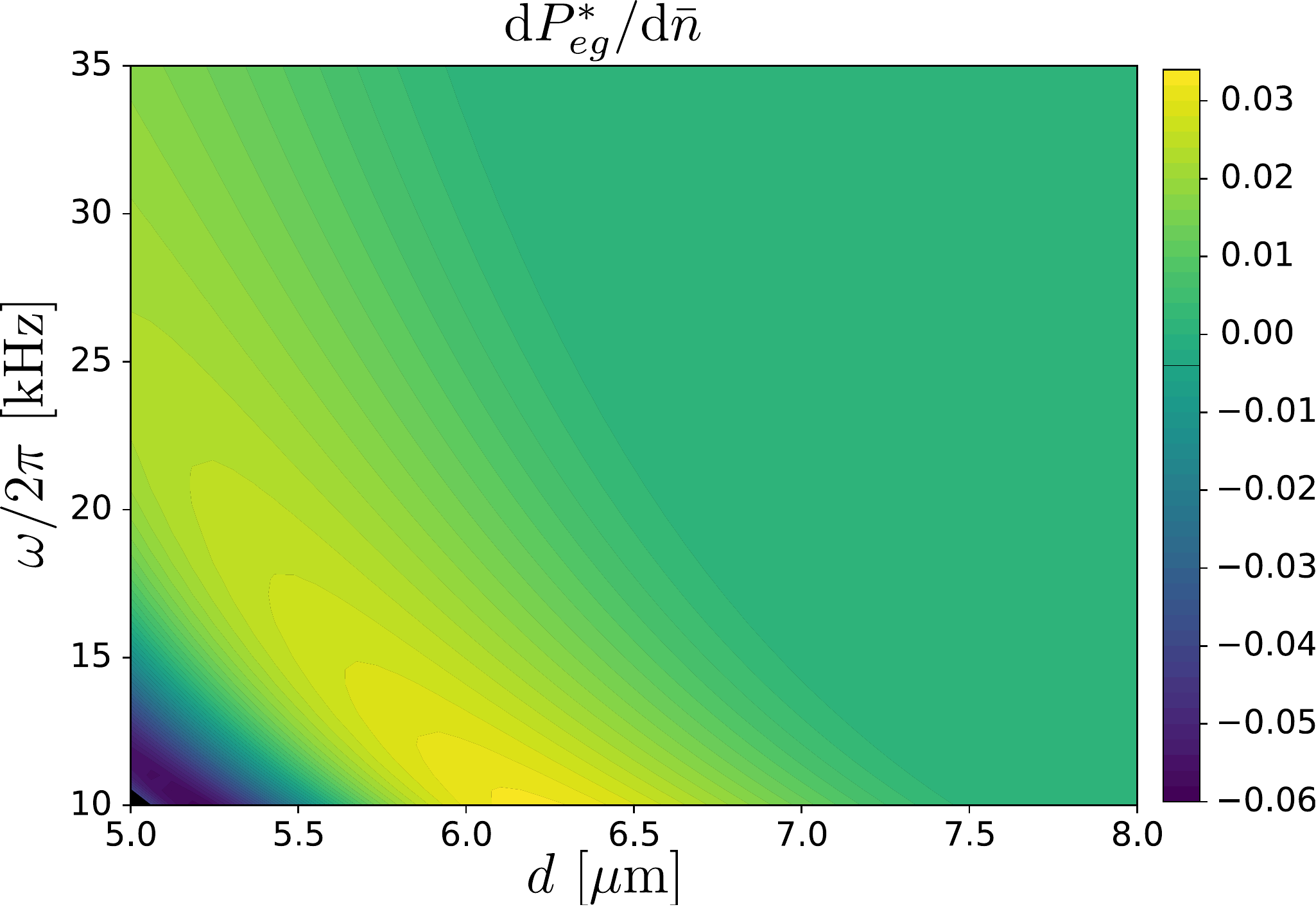}
\caption{Maps of the sensitivity~\eqref{eq:sensitivity} with $\bar{n}=1$ as a function of experimental control parameters $d$ and $\omega$ with other parameters fixed according to \textbf{THM}.}%
\label{fig:thermometrysensitivity}%
\end{figure}

The sensitivity discussion can be made more precise by computing $\dd{P_{eg}^*}/{\dd T}$ or, equivalently, $\dd{P_{eg}^*}/{\dd\bar{n}}$. Using the linear regime expressions, we get
\begin{align}
\label{eq:sensitivity}
\dv{P_{eg}^*}{\bar{n}} =&\; 4\alpha^2(1-\cos\varphi_R)\cos(4\alpha\alpha_z\sin\varphi_R)  \\
&\quad\times  e^{-4\alpha^2(1-\cos\varphi_R)(2\bar{n}+1)}\ ,
\nonumber
\end{align}
where $\varphi_R = \omega T_R$. Figure \ref{fig:thermometrysensitivity} shows a colormap plot of $\dd{P_{eg}^*}/{\dd\bar{n}}$ for $\bar{n}=1$ as a function of the only two remaining independent parameters, the interatomic distance $d$ and the trap frequency $\omega/2\pi$. The chosen parameter region ($d>\SI{5}{\micro\meter}$ and $\omega/2\pi > \SI{10}{\kilo\hertz}$) corresponds to a stability region of the traps (see Appendix~\ref{app:stability}), except in the zone where $\dv{P_{eg}^*}{\bar{n}}<0$. The \textbf{THM} parameter set is clearly close to the wide optimum found here, with about a 3\%/phonon maximum. Note that the sensitivity range of temperatures can be tuned by varying the experimental parameters  $d$ and $\omega$. For instance, for higher $\omega$ values, the sensitivity is lower but the measurement range can be extended to accordingly higher temperatures.

The precise measurement of the Rabi signal thus provides a sensitive thermometry for the initial atomic motion in the traps. It is most sensitive for rather small oscillation frequencies and for small average phonon numbers, conditions for which other methods (thermal expansion after releasing the trap, observation of motional sidebands...) are not readily applicable. The short involved timescale (for the \textbf{THM} set of parameters, $T_R/2=\SI{23.5}{\micro\second}$ only) makes the method also useful for simulators based on ordinary, laser-accessible short-lived Rydberg states. It could provide a useful tool for the diagnostic of Rydberg quantum simulators.

\section{Long-term spin exchange: temperature and non-linearity}

\begin{figure*}[t]%
\centering
\includegraphics[width=0.95\textwidth,clip]{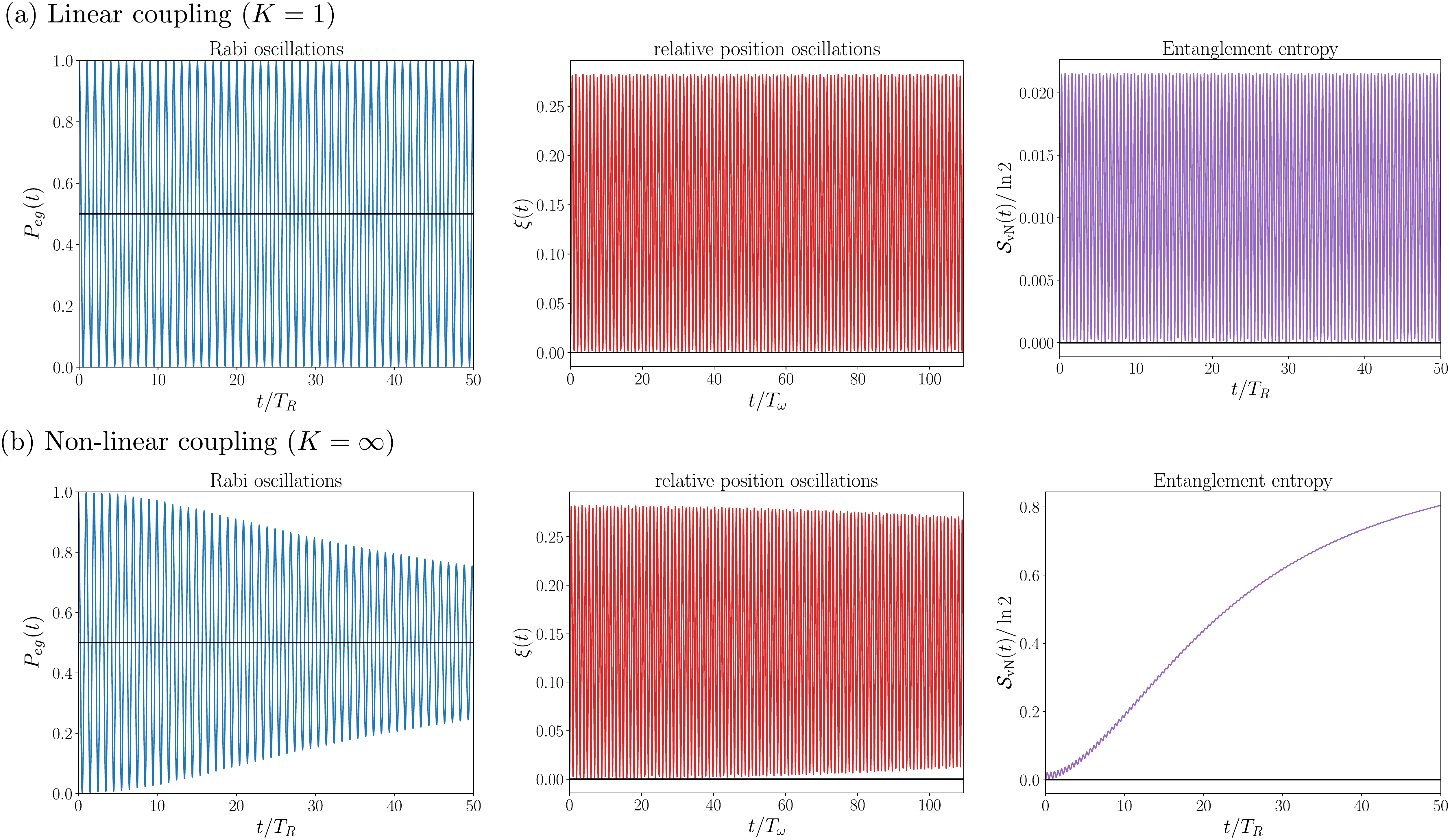}
\caption{Examples of Rabi oscillations for the experimental parameters of the \textbf{\XXZ} set in Table~\ref{tab:vdw}, with an initial temperature of $T=5\;\mu$K. The initial atomic state is $\ket{e,g}$. The linear coupling plots [upper panel (a)] correspond to $K=1$ while non-linear numerics [lower panel (b)] are performed with $K=\infty$. In both panels, we show first the Rabi oscillation signal $P_{eg}(t)$, the oscillations of the relative position $\xi(t)$ and the spin-motion entanglement entropy $\mathcal{S}_{\text{vN}}(t)$.}
\label{fig:XXZ}
\end{figure*}

We now come back to the original spin simulator of  Ref.~\onlinecite{NguyenQuantumSimulationCircular2018}. One of the key features of the circular-state quantum simulator is, that it makes, in principle, very long simulation times accessible. It is thus important to scrutinize the spin-motion behavior over longer timescales than in the previous sections. For a simple case, we focus on the spin-exchange Rabi oscillation for the \textbf{\XXZ} set of parameters given in Tab.~\ref{tab:vdw}. With a $\Delta n=2$ configuration and a large trapping frequency ($\omega/2\pi=\SI{50}{\kilo\hertz}$, $\omega/|J|=8.7$), this set of parameters is likely to minimize the influence of atomic motion. We will see that it is indeed the case at very low initial atomic temperatures, but that non-linearities of the interaction potential  (which can be treated perturbatively since $g$ is small) play a key role at a finite temperature.

The main results of this section are presented in Fig.~\ref{fig:XXZ}. In the upper panel, we show the results of the linear approximation ($K=1$) for the Rabi oscillation signal, the relative position oscillations and the entanglement entropy. The temperature is here chosen to be $\SI{5}{\micro\kelvin}$, but the linear regime results are almost temperature-independent in this range. Not surprisingly, we obtain a nearly unperturbed spin exchange Rabi oscillation signal over 50 periods (it remains unperturbed forever), with a negligible entropy (note the scale of the last frame).  

The exact Rabi oscillation signal including the spin-motion Hamiltonian to all orders ($K=\infty$) is unperturbed, too, over the first 50 Rabi periods, when assuming a zero initial temperature. However, as shown in the lower panels of Fig.~\ref{fig:XXZ}, the contrast of the Rabi oscillation decreases quite rapidly when the initial temperature is set to  $\SI{5}{\micro\kelvin}$. We analyze in details this effect to show that it is not induced by decoherence, as might be feared, but rather due to a gradual dephasing of the Rabi oscillations corresponding to different initial phonon numbers. We recall that we are considering a closed-system model. We will see that the Rabi oscillation signal exhibits periodic revivals with a full contrast, showing that this effect does not hamper the perspectives for a quantum simulator. In order to explore this dynamics analytically, we use the second-order expansion of the interaction potential $U(\hat{\xi})$ given by Eq.~\eqref{eq:expansion2}.

\subsection{Dynamics to second order}

Starting from \eqref{eq:xg} and rewriting it in dimensionless variables to second order in $\eta$, we get
\begin{align}
\ddot{\xi} + \omega^2 (\xi - \bar{\xi}) = 
-f\omega\eta^2\big(4J\expval*{\hat{\xi}\hSigma_X} -2\Delta {\xi}\big)\;.
\end{align}
Thus, we need to evaluate $\expval*{\hat{\xi}\hSigma_X}$. Since the pure state evolution from \eqref{eq:psit} generalizes to
\begin{equation}
\ket{\Psi(t)} = c_+\ket{\psi_+(x,t)}\ket{+X} + c_- \ket{\psi_-(x,t)}\ket{-X}\ ,
\end{equation}
one gets
\begin{align}
\xi(t) = \expval*{\hat{\xi}} & = \abs{c_+}^2 \xi_+(t) + \abs{c_-}^2 \xi_-(t)
\label{eq:xit}\\
\expval*{\hat{\xi}\hSigma_X} & = \abs{c_+}^2 \xi_+(t) - \abs{c_-}^2 \xi_-(t)\ ,
\label{eq:xiSigmaxt}
\end{align}
where $\xi_\pm(t)$ are the two mean trajectories in each single-particle potential $V_\pm(\hat{x})$:
\begin{equation}
\xi_\pm(t) = \expval*{\hat{\xi}}{\psi_\pm(x,t)}\;.
\end{equation}

\begin{figure}[t]%
\centering
\includegraphics[width=0.85\columnwidth,clip]{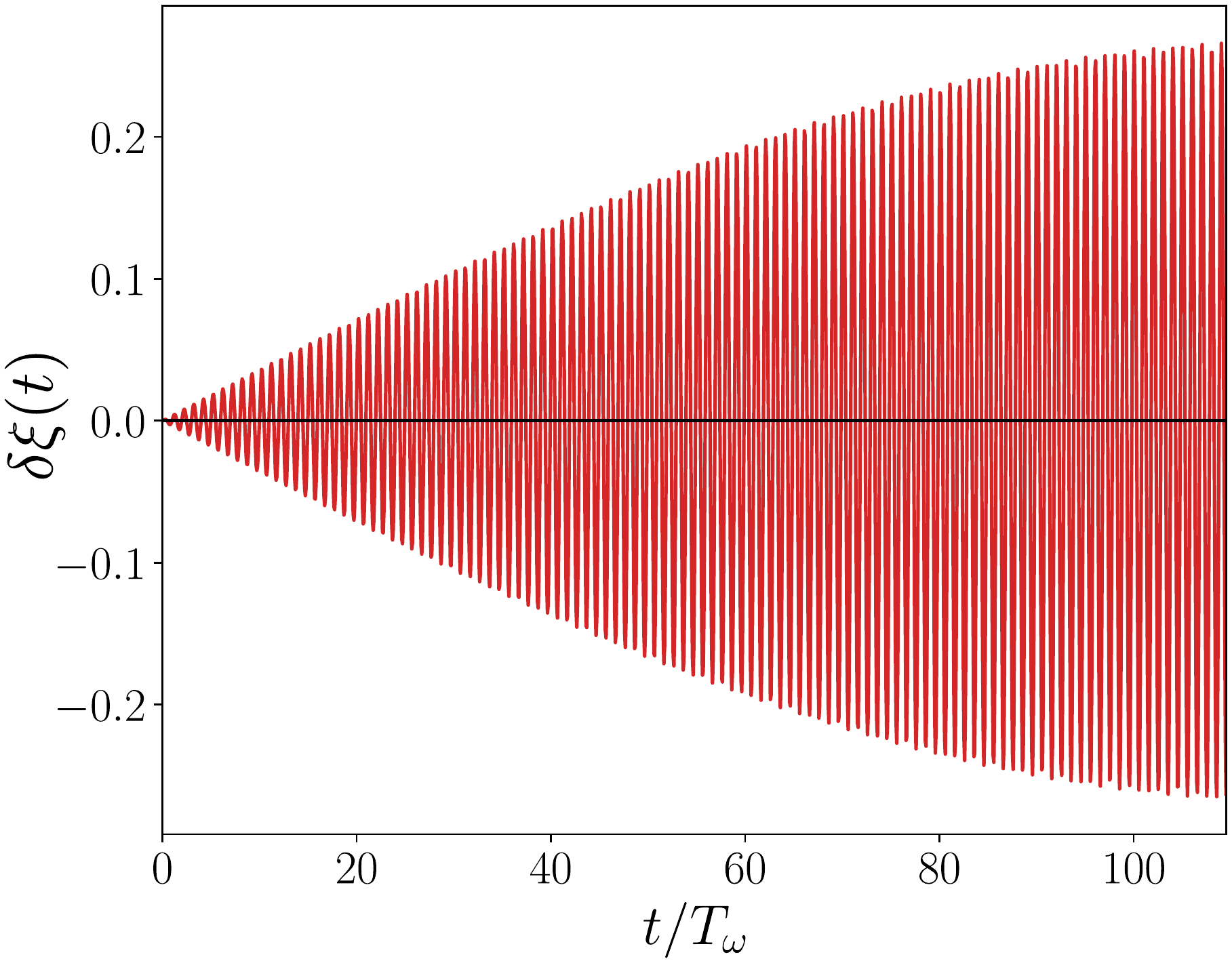}
\caption{Difference $\delta \xi(t)$ between results for linear $K=1$ and non-linear $K=\infty$ coupling on the relative position $\xi(t)$ with the parameters of Fig.~\ref{fig:XXZ}.}%
\label{fig:non-linear}
\end{figure}

Since one can apply Ehrenfest theorem for these two single-particle time evolutions, we get the following equations of motion up to order two in $\eta$:
\begin{equation}
\ddot{\xi}_\pm(t) + \omega_\pm^2 [\xi_\pm(t) - \bar{\xi}_\pm] = 0\ ,
\end{equation}
where
\begin{align}
\omega_\pm &= \omega\sqrt{1+2f\eta(\pm\alpha-\alpha_z)} \\
\bar{\xi}_\pm &= \frac{2(\pm\alpha-\alpha_z)}{1+2f\eta(\pm\alpha-\alpha_z)}\ .
\end{align}
Since the evolution equation in \eqref{eq:xg} involves the derivative of $U(\hat{\xi})$, the dependence in $\hat{\xi}$ is still linear at second order in $\eta$, which makes the calculation tractable.
The trajectories thus remain two harmonic oscillations that start from the same initial condition $\xi_0$. Assuming no initial velocity, we have:
\begin{equation}
\xi_\pm(t) = \bar{\xi}_\pm + (\xi_0-\bar{\xi}_\pm)\cos(\omega_\pm t)\ .
\end{equation}
The effect of non-linearity is to displace the mean position and amplitude of the oscillations but most importantly to lift the degeneracy of their frequencies and introduce beating when combining the two trajectories $\xi_{\pm}$ in observables such as the relative distance $\xi(t)$ in Eq.~\eqref{eq:xit}. An example of such beating with the \textbf{\XXZ} set of parameters is given in Fig.~\ref{fig:non-linear} by plotting the difference between the position computed with all non-linear terms ($K=\infty$) and the position computed at linear coupling ($K=1$), i.e..  $\delta\xi(t) = \xi^{(K=\infty)}(t)-\xi^{(K=1)}(t)$.

\subsection{Rabi oscillations contrast loss by dephasing}

\begin{figure}[t]%
\centering
\includegraphics[width=\columnwidth,clip]{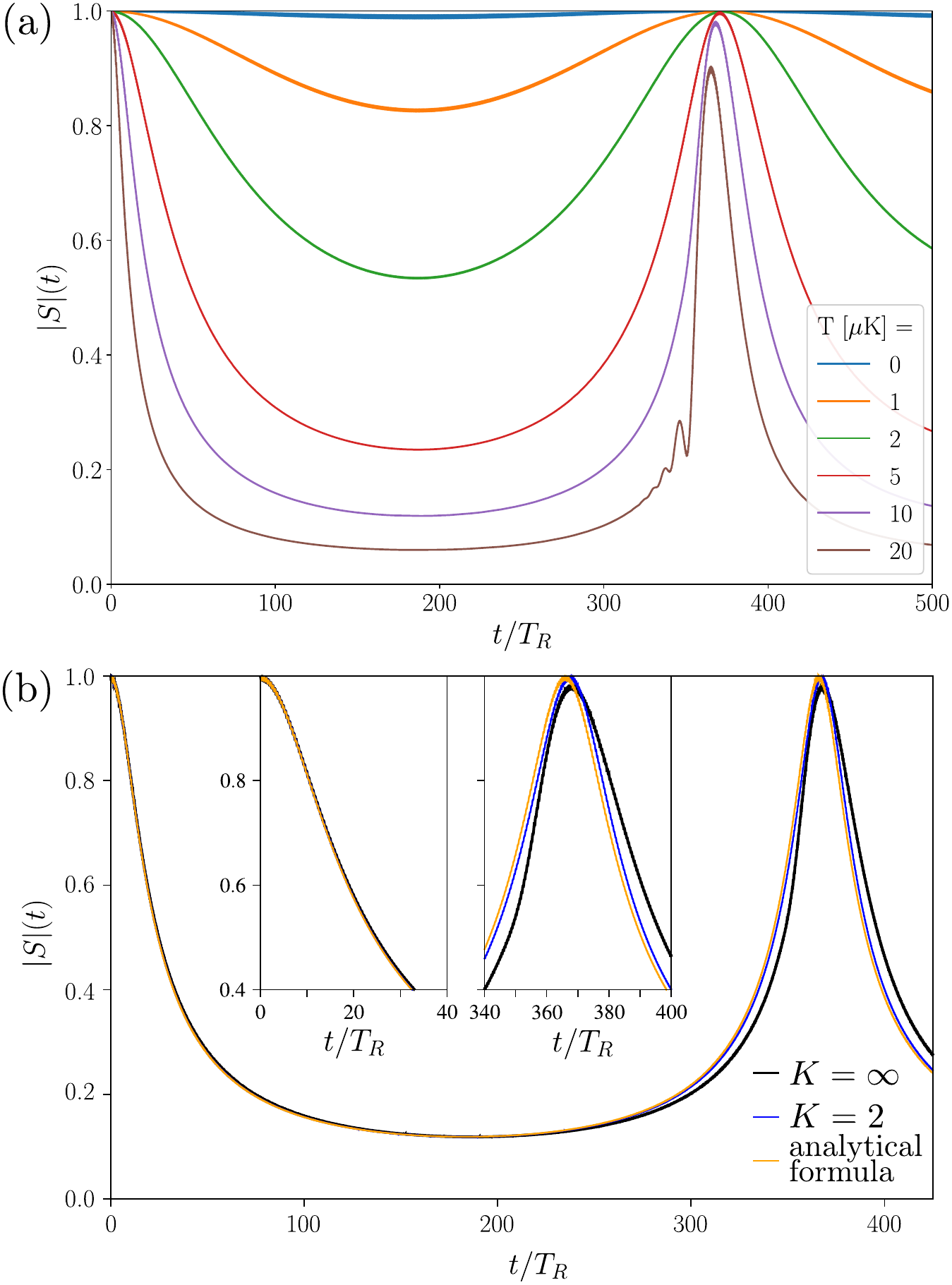}
\caption{(a) Effect of temperature on the contrast of Rabi oscillations in presence of non-linear effects with the parameters of Fig.~\ref{fig:XXZ}. (b) Comparison between the $K=2$ and $K=\infty$ results for $T=10\mu$K with equation \eqref{eq:St2}. Insets are zooms at short times and close to the revival peak.}%
\label{fig:decay}
\end{figure}

In Fig.~\ref{fig:XXZ}, we observe that the main effect of non-linearities is to induce an appreciable contrast loss. In Fig.~\ref{fig:decay}(a), we show the envelope of the Rabi oscillations obtained by computing
\begin{align}
\abs{S}(t) = \sqrt{s_Y^2(t) + s_Z^2(t)}\ ,
\end{align}
with the same parameters, but extending the calculation up to time $t=500\, T_R$. We observe an almost perfect revival of the Rabi oscillations around $370\,T_R$ and a dependence on temperature much stronger with non-linear effects than in the linear prediction of Fig.~\ref{fig:XXZ}.

We interpret this effect as resulting from the gradual dephasing of the Rabi oscillations corresponding to the different initial phonon numbers $n$ present in the initial thermal motional state.
Indeed, we can treat the second order non-linear term as a perturbation from the linear-coupling Hamiltonian, $\hat{H}_0$, given in~\eqref{eq:H1}, by writing the Hamiltonian as
\begin{equation}
\hat{H} = \hat{H}_{nl} \equiv \hat{H}_0 + \eta^2\frac{f}{2}\hat{\xi}^2 (2J\hSigma_X - \Delta)\ .
\end{equation}
Its eigenstates are the displaced Fock states $\hat{D}(\alpha\hSigma_X-\alpha_z)\ket{n,s}\, (s = \pm)$.
The energies to first order in perturbation theory read
\begin{align}
E(n,s) &= E_0(n,s)  + \eta\frac{f}{2}\omega\big\{
\alpha[(2n+1) + 4\alpha^2 + 12\alpha_z^2]s \nonumber\\
&\qquad -\alpha_z[(2n+1) + 4\alpha_z^2 + 12\alpha^2] \big\}\ ,
\end{align}
where the $E_0(n,s)$ are given by Eq.~\eqref{eq:Ens}.
We thus see that the Rabi frequency now depends linearly on the phonon number $n$:
\begin{equation}
\Omega \rightarrow \Omega' + \Omega_d\, n\ ,
\end{equation}
where
\begin{align}
\Omega' &= \Omega + 2Jf\eta^2(\alpha+4\alpha^3+12\alpha\alpha_z) \\
\Omega_d &= 4Jf\eta^2\ .
\end{align}
With the \textbf{XXZ}  set of parameters, $\alpha \ll 1$, and we can simplify the Rabi-frequency to $\Omega \rightarrow \Omega + \Omega_d\, n$.

At a finite initial temperature, the $S(t) = s_Z(t) + is_Y(t)$ function is the incoherent average
\begin{align}
S(t) = \sum_{n=0}^\infty p_n S_n(t)
\label{eq:Ssum}
\end{align}
of the time evolved $S_n(t)$ issued from Fock state $n$, with the Boltzmann weights 
\begin{equation}
p_n = \frac{1}{1+\bar{n}}  \left(\frac{\bar{n}}{1+\bar{n}}\right)^n\;.
\end{equation}
As an ansatz for the $S_n(t)$ functions, we use the linear coupling result given by Eq. \eqref{eq:fock} extended to an arbitrary spin initial state:
\begin{equation}
S_n(t) = 2c_-c_+^* e^{-2\abs{\gamma(t)}^2} L_n(\abs{2\gamma(t)}^2) e^{i[\Omega t - \Theta(t)]}e^{i\Omega_d n t}\,,
\end{equation}
in which second-order terms only introduce a Rabi-frequency change, reflected by the last term  involving $\Omega_d$.
With this simple assumption, the series \eqref{eq:Ssum} is carried out using the generating function of the Laguerre polynomials, so that
\begin{align}
S(t) &= 2c_-c_+^* C(t)
\frac{e^{i[\Omega t - \Theta(t)]}}{1+\bar{n}(1-e^{i\Omega_d t})}\nonumber \\
& \qquad \times \exp\left\{\abs{2\gamma(t)}^2 \bar{n}
\frac{(1+\bar{n})(1-e^{i\Omega_d t})}{1+\bar{n}(1-e^{i\Omega_d t})}
  \right\}\ ,
 \label{eq:St2}
\end{align}
with $C(t)$ given by \eqref{eq:Ct}. More practically, starting from $\ket{e,g}$, the Rabi signal can be derived from 
\begin{align}
s_Z(t) &= C(t)\times \nonumber\\
&\frac{(1 +\bar{n})\cos(\Omega t- \Theta(t))-\bar{n}\cos(\Omega t - \Theta(t)+\Omega_d t)}{1+2\bar{n}(\bar{n}+1)(1-\cos(\Omega_d t))}\ .
\end{align}
Since $\Omega_d\ll\Omega$, this expression, and hence the Rabi signal  have a quasi-period $T_d=2\pi/\Omega_d$. It explains the gradual initial contrast loss and the nearly full Rabi-contrast revival observed in  Fig.~\ref{fig:decay}(a). 

At short times with respect to $T_d$, one gets a quadratic decay for the envelope
$\abs{S(t)} \simeq C(t) [1 - (t/\tau_d)^2]$, from the denominator in~\eqref{eq:St2}, with
\begin{equation}
\tau_d = 1/(\Omega_d\sqrt{\bar{n}(\bar{n}+1)})
\end{equation}
being a typical dephasing time. 

The minimum $C_\text{min}$ of the contrast occurs when $\Omega_d t = \pi$. Since $C(t)$ oscillates much faster than the revival signal, a good estimate of the minimum contrast  is given by
\begin{equation}
C_\text{min} \simeq \frac{e^{-8(1+2\bar{n})\alpha^2}}{1+2\bar{n}}\ .
\end{equation}
The dominant term at weak coupling ($\alpha\ll 1$) is the denominator, which, remarkably, does not depend on the coupling in this regime.

The contrast loss, due to the gradual dephasing between the various Fock states contributions, and the revivals obviously require an initial dispersion of the phonon number and are only predicted when including the second-order terms in the evolution.  We have considered here an initial thermal state, but the periodic collapses and revivals of $P_{eg}(t)$ would be observed for any initial state (other than a single Fock state), including of course coherent states.

In Fig.~\ref{fig:decay}(b), we compare the analytic formula at second order for $T=\SI{10}{\micro\kelvin}$ with numerical results at second order  ($K=2$) and at all orders ($K=\infty$). Remarkably, for this set of parameters, the second order formula, despite it involves further approximations, reproduces almost exactly the $K=2$ numerical result. The main effect of the higher-order non-linear terms is to deform the revival toward a shorter time (larger $\Omega_d$) in an asymmetric manner.

For the experimental parameters \textbf{\XXZ} considered in this section, the expected long-term dynamics of the spin exchange is thus obtained only when the initial atomic temperature is small at the scale of the temperature $T_h$ corresponding to the trap frequency. This is a strong requirement, even if sideband-cooling mechanisms can be used in principle to fulfill it. Nevertheless, the loss of contrast results from a mere dephasing of different Rabi frequencies, that could, in principle, be limited by echo techniques. 

\section{Conclusion}

We have explored in this paper various aspects of the coupling between spin exchange and motion in the context of spin-array quantum simulator based on interacting, trapped circular Rydberg atoms~\cite{NguyenQuantumSimulationCircular2018}, with quite realistic experimental parameters. We have shown that, in the tight trap regime, this coupling can be made small and that long-term spin-exchange Rabi oscillations are observed. Nevertheless, we have unveiled the essential role of a finite atomic temperature and of the non-linear behavior of the spin-motion interaction and shown that unperturbed simulations over very long timescales would require either echo methods to get rid of the Rabi oscillations dephasing or a careful cooling of the atoms in their vibrational ground state.

It would be thus important to assess the motional temperature in a low-phonon number regime, where standard methods might be of little use. We have shown that the spin exchange itself provides a solution. By relaxing the trapping potential, the spin-exchange Rabi signal can be made utterly sensitive to the atomic temperature, even on very short time scales. This thermometry method could also be useful in other situations, particularly for quantum simulators based on low angular momentum Rydberg atoms.

In the limit of an extremely strong coupling between the spin exchange and motion, obtained only for a resonant, first order dipole-dipole interaction between the atoms, we have shown that spin and motion become fully entangled, resulting in interesting motional states. The involved quantum superpositions of different coherent amplitudes, a Schr\"odinger-cat situation, is also realistically observable in present experiments~\cite{Mehaignerie2023}.

Obviously, this paper is only a first step in the exploration of a very rich situation. We have considered here only the spin-exchange dynamics to unveil the main mechanisms, but it would be straightforward, at least numerically, to extend to the full \XXZ\ dynamics, that is obtained by dressing the atomic transition with a nearly resonant microwave field. The extension to larger chains is also an interesting objective. How, for instance, would the propagation of a single excitation over a chain of atoms be modified by the coupling with the atomic motion? How does this experimentally achievable situation compares with, for instance, the spin-charge decoupling models? Obviously, the exact numerical approach will face severe problems, even for a few atoms.

\section{Acknowledgements}
This work has been supported by the European Union FET-Flag project n\textdegree 817482 (PASQUANS), ERC Advanced grant n\textdegree\ 786919 (TRENSCRYBE) and QuantERA ERA-NET (ERYQSENS, ANR-18-QUAN-0009-04), by the Region Ile-de-France in the framework of DIM SIRTEQ and by the ANR (TRYAQS, ANR-16-CE30-0026).

\appendix

\section{Considerations on the stability of effective potentials}
\label{app:stability}

This section briefly tackles situations when a strong coupling between spin and motion induces non-perturbative behaviors. This regime is undesirable experimentally and the goal of this section is to get quantitative estimates of the coupling required to enter this regime and some typical qualitative behaviors that one can observe for such parameters.

\subsection{Stability of spin-dependent potentials}

\begin{figure}[t]
\centering
\includegraphics[width=\columnwidth,clip]{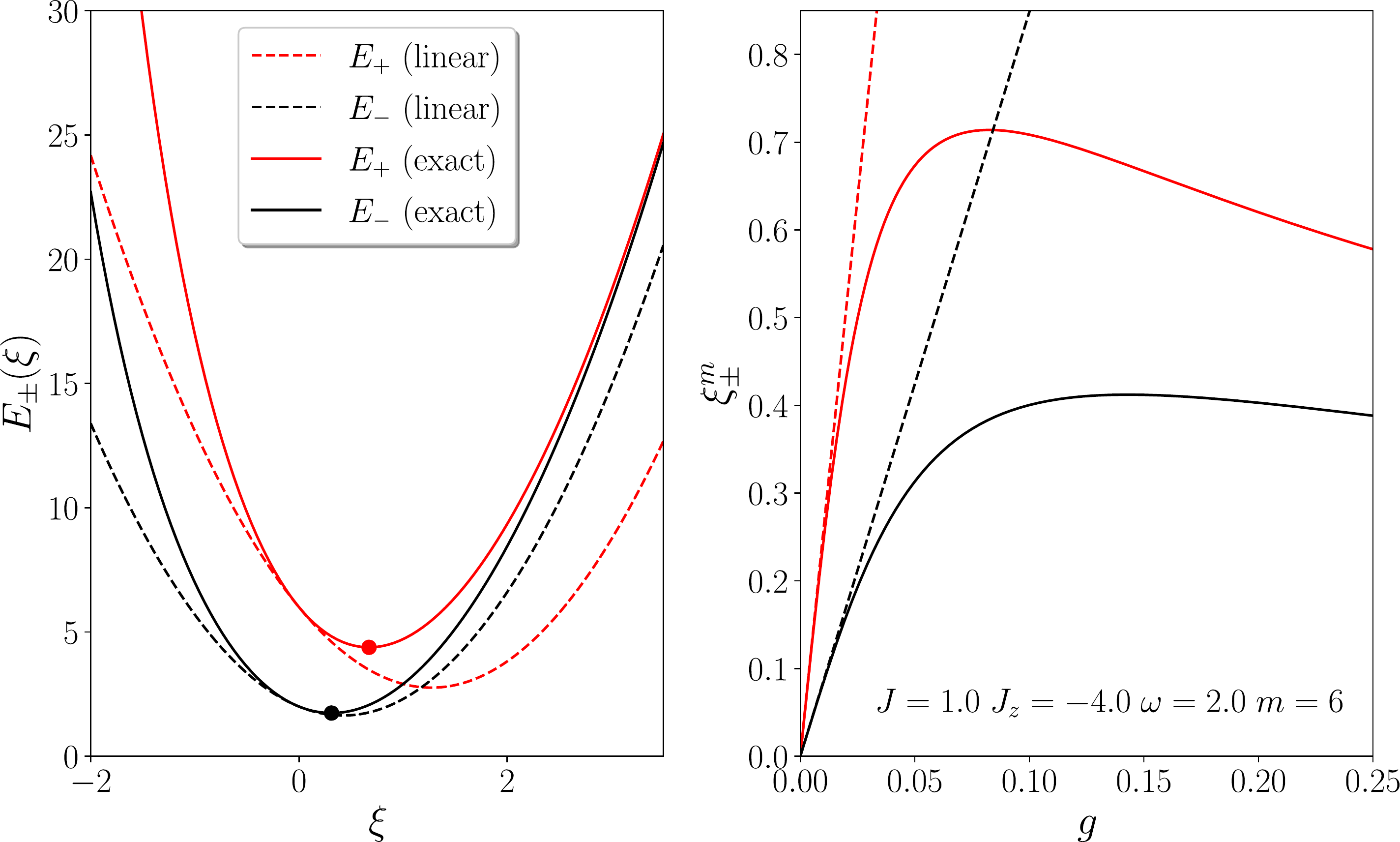}
\caption{(a) Example when the $V_{\pm}$ are both stable. (b) Evolution of the minima of these potentials with coupling $g$. Parameters are chosen for illustration purpose and do not correspond to experimentally realistic ones.}
\label{fig:stable}
\end{figure} 

On Fig.~\ref{fig:exppotential}, we see that the experimental trapping potentials cannot be approximated by a quadratic potential if the $x$-variable takes too large values. At large temperatures or couplings, while the system explores large values of $x$, this is one possible limitation of the previous description.
In this section, we wish to discuss another limitation that is intrinsic to Hamiltonian~\eqref{eq:Ham}, i.e., with harmonic traps, and that is relevant at large coupling $g$ when non-linear effects have to be taken into account.
Indeed, in the linearized version of the Hamiltonian in the previous sections, 
the two effective wells of \eqref{eq:Vpm_linear} remain parabola that are merely shifted in energy and position. However, using the dimensionless variable $\xi = x/x_0$, one rewrites the spin-dependent potentials as
\begin{equation}
\label{eq:Vpot}
\frac{V_{\pm}(\xi)}{\hbar} = \frac{\omega}{4}\xi^2 +
\frac{\pm 2J-\Delta}{(1+g\xi)^{m}}\ ,
\end{equation}
with $\xi \in [-1/g,+\infty]$. 
In this form, one sees that for $\xi \to -1/g$, the second term diverges. Whether the potential diverges to $\pm\infty$ depends on the sign of the numerator. If the two prefactors are positive, i.e., when\begin{equation}
\label{eq:stability}
 2\abs{J} < -\Delta \;,\quad (\text{strong stability condition})
\end{equation}
stability of both potentials around their minima is ensured, whatever the coupling strength $g$. Clearly, if $\Delta \geq 0$, this condition is never fulfilled. Thus, this strong stability requires $\Delta <0$ and $J$ not too large, since $2\abs{J} < \abs{\Delta}$.
The extrema of \eqref{eq:Vpot} are solutions of the equation
\begin{equation}
\label{eq:extrema}
\xi  (1+g\xi)^{m+1} = g\frac{2m}{\omega}(\pm 2J-\Delta)\; ,
\end{equation}
for which the two minima $\xi_0^{\pm} \simeq x_0^{\pm}/x_0 $ at weak-coupling $g$ are those given in Eq.~\eqref{eq:x0pm}.
In the strong-stability condition case, these minima are positive and go through an intermediate maximum before decaying towards zero as
\begin{equation}
\xi_0^{\pm} \simeq \left(2m\frac{|\Delta\mp 2J|}{\omega} \right)^{1/(m+2)}
\; \frac{1}{g^{m/(m+2)}}
\end{equation}
at large coupling $g$. This behavior is illustrated in Fig.~\ref{fig:stable}. In such case, the overall qualitative interfering paths picture remains correct, with trajectories that are deformed but nothing dramatic happens.

\begin{figure}[tb]
\centering
\includegraphics[width=\columnwidth,clip]{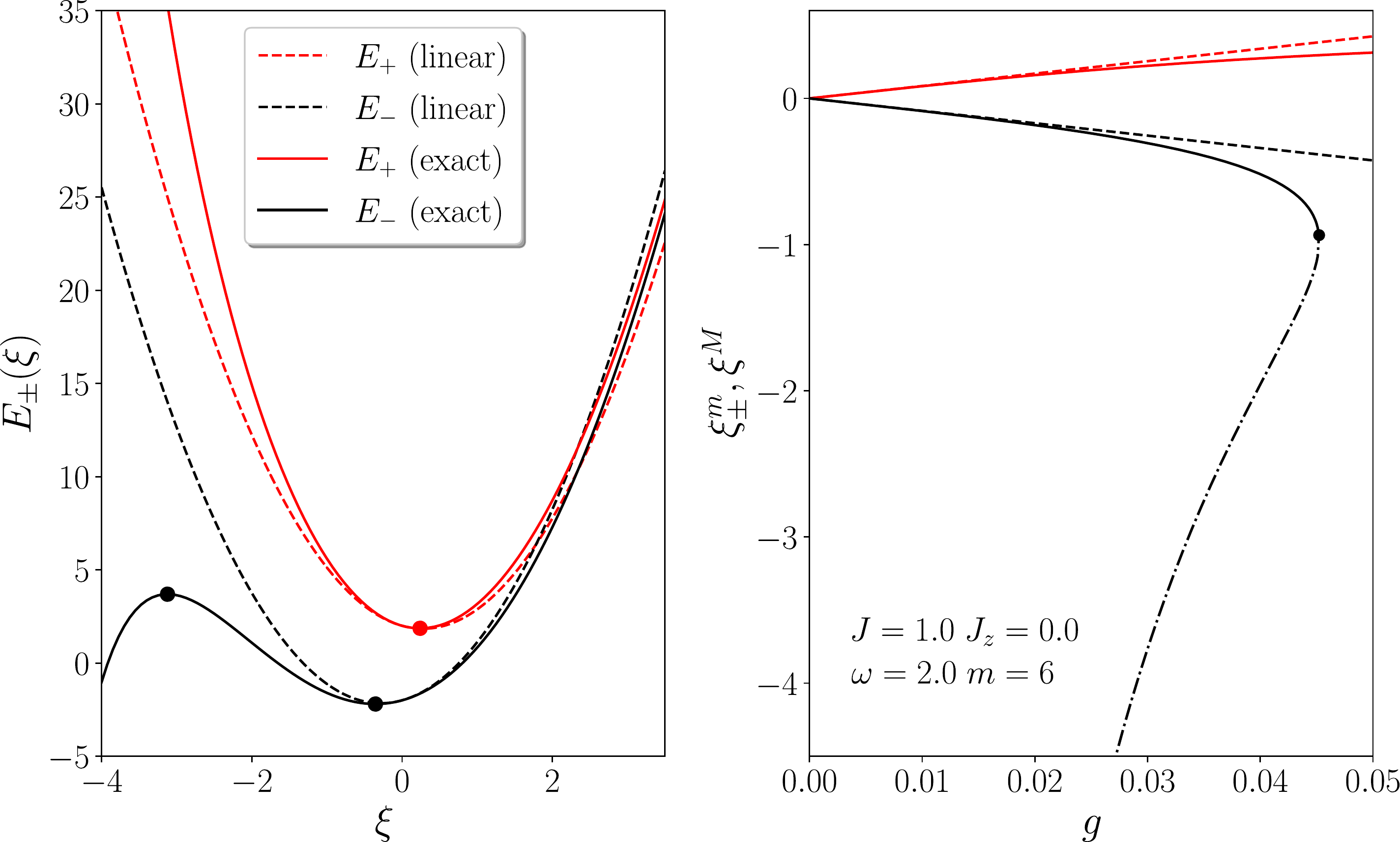}
\caption{(a) Same as Fig.~\ref{fig:stable} but in the case where one of the potential is metastable. (b) Evolution of the minima (full lines) and maximum (dashed lines) as a function of the coupling $g$. }
\label{fig:unstable}
\end{figure}

When \eqref{eq:stability} is not satisfied, there typically exists at least one potential with a negative divergence branch as illustrated in Fig.~\ref{fig:unstable} for a metastable $V_-$ potential. 
Metastability is associated to the existence of an intermediate maximum $\xi_M$ solution to \eqref{eq:extrema}, that is necessarily negative.
At weak coupling $g$, we directly see in \eqref{eq:extrema} that it scales as
\begin{equation}
\xi_M \simeq -1/g\;.
\end{equation}

When increasing $g$, the maximum eventually merges with the minimum of the same branch at a critical coupling $g_c$ beyond which the minimum disappears, corresponding to a breakdown of the laser-trap stability. Before that, tunneling through the metastable potential could already become significant. 
Consequently, the metastability region is given by the conditions 
\begin{equation}
2\abs{J} > -\Delta \text{ and } g < g_c \;,\;\text{ (metastability condition)}\ .
\end{equation}
The value of the critical coupling reads
\begin{equation}
\label{eq:gc}
g_{c,\pm}^{(m)} = \frac{1}{\sqrt{2m}} \frac{(m+1)^{(m+1)/2}}{(m+2)^{(m+2)/2}}\sqrt{\frac{\omega}{\vert{\Delta \mp 2J}\vert}}\; . 
\end{equation}
The associated value of the critical dimensionless relative distance is
\begin{equation}
\xi_c = -\frac{1}{(m+2)g_c} \;.
\end{equation}
More explicitly, for $m=3$ (for which we can neglect $\Delta$) and $m=6$, Eq.~\eqref{eq:gc} gives
\begin{align}
g_{c,\pm}^{(3)} &\approx 0.116847 \sqrt{\frac{\omega}{\vert{2J}\vert}}\;,\\
g_{c,\pm}^{(6)} &\approx 0.063958 \sqrt{\frac{\omega}{\vert{\Delta \mp 2J}\vert}}\;.
\end{align}
These values may become small when $\omega < {\vert{\Delta \mp 2J}\vert}$. We observe that only at sufficiently low trap frequencies $\nu$ and distance $d$, in which one reaches strong couplings $g$ one goes beyond the critical value $g_{c,-}$.

\begin{figure*}[t]%
\centering
\includegraphics[width=\textwidth,clip]{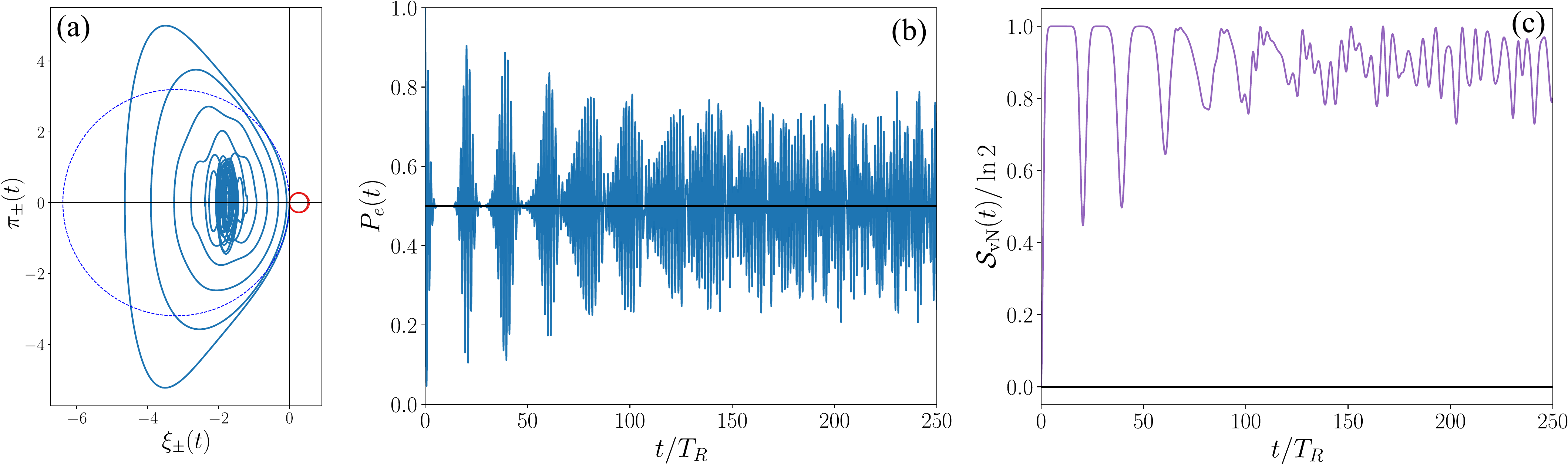}
\caption{Typical behavior in the  metastability regime for temperature $T=\SI{0}{\micro\kelvin}$  and parameters described in the main text. (a) $\xi_{\pm}$ phases space trajectories. (b) Rabi oscillations signal $P_{eg}(t)$. (c) Entanglement entropy $\mathcal{S}_{\text{vN}}(t)$.}%
\label{fig:panic}
\end{figure*}

\subsection{Typical behavior of observables}

In Fig.~\ref{fig:panic}, we give a typical example of behavior in the 
strong coupling regime. We choose a $\Delta n=2$ situation but starting with $n=50$ and with slightly different magnetic and electrical fields. The main parameters are $d=\SI{6}{\micro\meter}$ and $\omega=\SI{7}{\kilo\hertz}$ so that $\alpha \simeq -1.38$ and $g\simeq 0.021$ is large but with $g \lesssim g_{c,-}$ so that metastable trapping occurs.
In Fig.~\ref{fig:panic}(a), the trajectory in the metastable well (in blue) becomes strongly deformed and eventually oscillates in a confined region of the well. On the contrary, the stable well has an almost harmonic behavior with small oscillations close to the minimum. The Rabi oscillations of Fig.~\ref{fig:panic}(b) display deformed revivals after a few periods $T_R$ but eventually transform in a structure-less signal with many frequencies. These two aspects also reflect in the behavior of the entanglement entropy of Fig.~\ref{fig:panic}(c). 

\onecolumngrid

\section{Coherent state dictionary and calculation of observables in the linear coupling limit}
\label{app:formulas}

With the exact result \eqref{eq:rhooft}, one can derive all possible observables concerning the position and corelators between position and spin using standard formula for coherent states.
We briefly recall that traces of position observables are best computed in the coherent state phase space representation~\cite{GlauberCoherentincoherentstates1963}.
For instance, using the coherent state representation of the thermal density matrix, one writes
\begin{equation}
\Tr(\rho_\beta D(\gamma)) = \frac{1}{\pi\bar{n}}\int\dd^2\alpha\; e^{-\abs{\alpha}^2/\bar{n}}\bra{\alpha}{D(\gamma)}\ket{\alpha}
= e^{-\abs{\gamma}^2(\bar{n}+1/2)}\;.
\label{eq:B1}
\end{equation}
This generalizes to
\begin{equation}
\Tr(\rho_\beta\,\hat{\xi}\, D(\gamma)) = \frac{1}{\pi\bar{n}}\int\dd^2\alpha\; e^{-\abs{\alpha}^2/\bar{n}}\bra{\alpha}{(a+a^\dagger)D(\gamma)}\ket{\alpha}
= \left((\bar{n}+1)\gamma-\bar{n}\gamma^* \right)e^{-\abs{\gamma}^2(\bar{n}+1/2)}\;.
\label{eq:B2}
\end{equation}
Note that in formula \eqref{eq:B1} and \eqref{eq:B2}, $\gamma$ is a generic parameter and does not correspond to \eqref{eq:gammar}, on the contrary to the next formulas below.
As an application, we get the following results for the simplest observables involving the relative position:
\begin{align}
n(t) &= \bar{n} + \abs{c_+}^2\abs{\gamma_0(t) + \gamma(t)}^2
+ \abs{c_-}^2\abs{\gamma_0(t) - \gamma(t)}^2\\
\sigma^2(t) &=\bar{\sigma}^2 + (1+2\bar{n})\left[\abs{c_+}^2\abs{\gamma_0(t) + \gamma(t)}^2
+ \abs{c_-}^2\abs{\gamma_0(t) - \gamma(t)}^2 \right]\\
\xi(t) &= 2\left[\abs{c_+}^2\Re\left(\gamma_0(t)+\gamma(t)\right) + \abs{c_-}^2 \Re(\gamma_0(t)-\gamma(t))\right]\\
\expval*{\hat{\xi}^2}(t) &= 1+2\bar{n} + 
4\left[\abs{c_+}^2\Re(\gamma_0(t)+\gamma(t))^2
+ \abs{c_-}^2\Re(\gamma_0(t)-\gamma(t))^2\right]\\
\expval*{\hat{\xi}\Sigma_X}(t) & = 2\left[\abs{c_+}^2\Re\left(\gamma_0(t)+\gamma(t)\right) - \abs{c_-}^2 \Re(\gamma_0(t)-\gamma(t))\right]\\
\expval*{\hat{\xi}^2\Sigma_X}(t) &= (\abs{c_+}^2 - \abs{c_-}^2)(1+2\bar{n}) + 
4\left[\abs{c_+}^2\Re(\gamma_0(t)+\gamma(t))^2
- \abs{c_-}^2\Re(\gamma_0(t)-\gamma(t))^2\right]\\
\expval*{\hat{\xi}\Sigma_Y}(t) & = 
4C(t)\Im{c_-c_+^*e^{i\Omega t-i\Theta(t)}
\left(\Re(\gamma_0+\gamma)-\gamma(\bar{n}+1)+\gamma^*\bar{n}\right)}
\\
\expval*{\hat{\xi}\Sigma_Z}(t) & = 
4C(t)\Re{c_-c_+^*e^{i\Omega t-i\Theta(t)}
\left(\Re(\gamma_0+\gamma)-\gamma(\bar{n}+1)+\gamma^*\bar{n}\right)}\ ,
\end{align}
in which $\sigma^2 = \ev{\hat{n}^2}-\ev{\hat{n}}^2$ are number fluctuations in the position, with the thermal expectation $\bar{\sigma}^2=\bar{n}(\bar{n}+1) $.

\twocolumngrid

\end{document}